\documentclass[fleqn,usenatbib]{mnras}

\usepackage{newtxtext,newtxmath}

\usepackage[T1]{fontenc}

\DeclareRobustCommand{\VAN}[3]{#2}
\let\VANthebibliography\thebibliography
\def\thebibliography{\DeclareRobustCommand{\VAN}[3]{##3}\VANthebibliography}

\usepackage{graphicx}	
\usepackage{amsmath}	

\title[\textsc{jz-fmm}: GPU-native differentiable FMM]{\textsc{jz-fmm}: GPU-native differentiable N-body simulations with the Fast Multipole Method}

\author[J. Stücker and O. Foldal]{
Jens Stücker$^{1}$\thanks{E-mail: jens.stuecker@univie.ac.at} and
Oskar Foldal$^{1}$
\\
$^{1}$Department of Astrophysics, University of Vienna, T\"urkenschanzstraße 17, 1180 Vienna, Austria
}

\date{Accepted XXX. Received YYY; in original form ZZZ}

\pubyear{\the\year{}}

\begin{document}
\label{firstpage}
\pagerange{\pageref{firstpage}--\pageref{lastpage}}
\maketitle

\begin{abstract}
N-body simulations are an essential tool for modelling the evolution of gravitating systems. \emph{Differentiable} N-body simulations allow solving complicated reconstruction problems by connecting observations of evolved systems to simple informative priors on their initial conditions. While so far most reconstruction efforts focus on mildly non-linear large scale dynamics, the internal dynamics of galaxies and clusters should exhibit rich information about their formation. The exploration of reconstruction possibilities in this highly non-linear regime naturally requires highly performant differentiable N-body simulations with small scale forces. Here, we present \textsc{jz-fmm}, a GPU-native differentiable implementation of the fast multipole method for N-body simulations. We show that the GPU oriented design of tree structure, dual tree traversal and multipole translation operators improves performance over established CPU-native and hybrid N-body codes by notably more than an order of magnitude. Further, we show that gradients can be evaluated efficiently and accurately so that \textsc{jz-fmm} can be used as an essential building block in future field-level reconstruction efforts. We apply the new code to the reconstruction problem of a tidally stripped satellite to show that deeply non-linear problems can indeed be solved efficiently with simulation gradients, although appropriate care must be taken to navigate the complicated optimization landscape.
\end{abstract}

\begin{keywords}
gravitation -- methods: analytical -- methods: numerical -- galaxies: kinematics and dynamics -- cosmology: theory
\end{keywords}

\newcommand{\mat}[1]{\mathbf{#1}}
\renewcommand{\vec}[1]{\mathbf{#1}}



\section{Introduction}

N-body simulations are a central tool for studying the dynamical evolution of self-gravitating systems, from star clusters and galaxies to the large-scale structure of the Universe \citep[see reviews by][]{trenti_hut_2008,dehnen_read_2011,angulo_hahn_2022}. Their computational cost is commonly dominated by the gravitational force calculation. Due to the quadratic scaling of direct summation, a variety of approximate methods is employed across modern N-body codes. Particle-mesh (PM) methods calculate forces through efficient convolutions on a regular grid, but their force resolution is tied to the mesh spacing \citep{hockney_1965,eastwood_hockney_1974,List_2026}. Adaptive meshes increase the spatial resolution where it is needed and can use multigrid techniques to solve the resulting field equations efficiently \citep{brandt_1977,teyssier_2002}. Alternatively, tree-based hierarchical algorithms adaptively group distant source particles and approximate their combined gravitational field, as pioneered by early tree methods \citep{appel_1985,barnes_hut_1986}. Hybrid TreePM schemes combine the efficient long-range force calculation and periodic boundary conditions of PM methods with the higher spatial resolution of a hierarchical short-range solver \citep{bagla_2002,springel_2005}.

The fast multipole method (FMM) provides a particularly powerful hierarchical approach \citep{rokhlin_1985,greengard_rokhlin_1987}. Whereas the Barnes--Hut method summarizes distant source regions, the FMM introduces expansions around both source and target regions, allowing the interactions between entire groups of particles to be evaluated collectively. This two-sided approximation enables linear asymptotic complexity and offers systematic control over the force accuracy through the expansion order and separation criterion. The classical three-dimensional FMM represents the Newtonian or Coulomb kernel through spherical harmonics \citep{cheng_1999}, whereas Cartesian FMMs instead use Taylor tensors. While Cartesian FMMs generally require more coefficients at high order, their generality, simple translation operators and small computational prefactors can make them competitive at the modest expansion orders relevant to many applications, while traceless-tensor specializations can further reduce their storage and computational costs for harmonic kernels. They have consequently been adopted and studied in a wide range of implementations \citep{tausch_2003,dehnen_2002,engblom_2011,dehnen_2014,coles_bieri_2020}. In gravitational N-body systems, symmetric mutual cell--cell interactions additionally allow Cartesian formulations to satisfy Newton's third law by construction and thus conserve total momentum \citep{dehnen_2000,dehnen_2002,dehnen_2014}.
The FMM has since been adopted by many modern astrophysical codes including \textsc{pkdgrav3}, \textsc{gadget4} and \textsc{swift}, and has recently also been implemented as a Poisson solver in \textsc{ramses} \citep{potter_2017,springel_2021,schaller_2024,lee_teyssier_2026}. For periodic cosmological simulations, FMM--PM schemes additionally combine the FMM short-range solver with a mesh-based long-range force \citep{springel_2021,schaller_2024}.

At the same time, the increasing availability of GPUs has shifted scientific computing towards highly parallel accelerator architectures. Gravitational N-body calculations are natural candidates for GPU acceleration because they require a large number of similar interaction evaluations. Existing implementations range from GPU-resident hierarchical force solvers \citep[e.g.][]{bedorf_2012,miki_umemura_2017} to hybrid approaches in which tree construction, communication or other stages remain on the CPU \citep[e.g.][]{ogiya_2013,habib_2016,garrison_2021,potter_2017,ragagnin_2020,wang_meng_2021,ragagnin_2026}. Hierarchical methods nevertheless remain challenging to map efficiently to accelerators: tree construction and traversal involve irregular data structures, branching and communication, and these stages can become significant when only the interaction kernels are offloaded. A fully GPU-native approach, in which sorting, tree construction, interaction-list generation and force evaluation all remain on the accelerator, therefore still offers an opportunity to meaningfully improve the end-to-end performance of N-body simulations.

Beyond predicting the evolution of a known initial state, a growing class of applications asks the inverse question: which initial conditions or physical parameters could have produced an observed final system? Although gravitational dynamics are in principle reversible, observations generally provide incomplete phase-space information and hence do not uniquely determine the past. The solution space can, however, be restricted substantially by combining the dynamical model with informative priors. This principle underlies cosmological field-level inference, where gradient-based methods have been used to connect observed large-scale structure to probable initial density fields \citep[e.g.][]{jasche_wandelt_2013,wang_2014,modi_2018,jasche_lavaux_2019,mcalpine_2026,List_2026}. Existing differentiable N-body codes developed for this application focus on pure particle-mesh approaches whose force resolution is limited by the mesh spacing -- typically in the megaparsec range \citep[e.g.][]{modi_2021,li_2024,List_2026}.

Most applications of differentiable cosmological simulation have so far concentrated on large or mildly non-linear scales, where perturbative or particle-mesh descriptions are computationally effective. Multiresolution approaches have begun to extend field-level inference towards individual galaxy scales, enabling constrained reconstructions of the Milky Way--M31 system and its surrounding matter distribution \citep{wempe_2024,wempe_2026}. Extending such techniques further into the internal dynamics of galaxies could make it possible to incorporate information from tidal debris, satellite populations and orbital histories into dynamical inference, with potential applications to host potentials, satellite progenitors and assembly histories. This regime is, however, considerably more difficult: systems may evolve over several orbital times, become strongly non-linear or chaotic, and exhibit complicated degeneracies between their initial conditions. Efficient differentiable N-body solvers with accurate short-range forces are needed to investigate how much of this information can be recovered in practice.

In this work, we present \textsc{jz-fmm}, a GPU-native implementation of the FMM for fast and differentiable N-body simulations. The code is open source and available under the MIT licence.\footnote{\url{https://github.com/jstuecker/jzfmm}} The code has two principal objectives: to advance the performance of GPU-based gravitational force evaluations and to enable differentiation through adaptive FMM force evaluations.

To achieve optimal performance, the code is designed from the ground up around a GPU-native computation model. High-level control flow is implemented in the \textsc{JAX} numerical computing library while performance-critical operations use NVIDIA's \textsc{CUDA} programming platform through \textsc{JAX}'s foreign function interface (FFI). Central to this design is a bottom-up tree construction that enables efficient, coalesced memory access during dual tree traversal \citep{stuecker_2026}. In our benchmarks, \textsc{jz-fmm} surpasses the force-evaluation performance of established CPU based and hybrid N-body codes by notably more than an order of magnitude.

To achieve differentiability, we derive a custom vector-Jacobian product (VJP) rule that advects gradients through the FMM using a simple second FMM calculation with modified inputs, rather than applying automatic differentiation to the individual FMM operations. We verify the accuracy of the FMM and the gradient computation through a series of numerical tests.

Another differentiable FMM implementation, \textsc{jaxFMM}, was recently presented by \citet{kraft_2026}, using a pure JAX implementation and automatic differentiation. In the implementation described in that paper, interaction-list construction was not compatible with just-in-time (JIT) compilation, limiting its suitability for repeated evaluations with changing particle positions. Our approach combines a fully JIT-compatible force-evaluation pipeline, explicitly optimized CUDA kernels, and an analytically derived custom VJP rule for the FMM.

Finally, we apply \textsc{jz-fmm} to a simple satellite reconstruction problem. Starting only from the final particle distribution and a prior on the initial satellite structure, we infer the satellite's initial position and velocity with gradient descents through a differentiable N-body simulation. This proof-of-concept application illustrates both the opportunities and the challenges of extending dynamical reconstruction methods into the strongly non-linear regime.

\section{Numerical Framework}

\subsection{The N-body system}

N-body systems are governed by the Hamiltonian
\begin{align}
    H = \frac{1}{2} \sum_i m_i \vec{v}_i^2 + \frac{1}{2} \sum_{i \neq j} G m_i m_j g(\lVert \vec{x}_i - \vec{x}_j \rVert)
\end{align}
where $\vec{v}_i$ and $\vec{x}_i$ are the velocity and position vectors of particles respectively. The
second sum fundamentally evaluates the convolution of a density field made of discrete point masses with the gravitational potential kernel $g(r)$. For the pure N-body system
\begin{align}
    g(r) = - \frac{1}{r}
\end{align}
However, we will default here to simulations with softened interactions, using the Plummer potential \citep{plummer_1911, aarseth_1963},
\begin{align}
    g_{\textrm{plummer}}(r) = - \frac{1}{\sqrt{r^2 + \epsilon^2}}
\end{align}
where the softening $\epsilon$ is a free parameter. We use $g$ for the interaction kernel and $\phi$ for the potential field generated by the particles.

Our default integration scheme is the Drift-Kick-Drift (DKD) form of the leapfrog integrator \citep[e.g.][]{quinn_1997,springel_2005}
\begin{align}
    \vec{x}_i(t + \Delta t/2) &= \vec{x}_i(t) + \frac{1}{2} \vec{v}_i(t) \Delta t \\
    \vec{v}_i(t + \Delta t) &= \vec{v}_i(t) + \vec{F}_i(t + \Delta t/2) \Delta t \\
    \vec{x}_i(t + \Delta t) &= \vec{x}_i(t + \Delta t/2) + \frac{1}{2} \vec{v}_i(t + \Delta t) \Delta t \\
\end{align}
where the potential at particle $i$, excluding its self-contribution, and the accelerations $\vec{F}_i$ (or loosely: 'forces') are given by
\begin{align}
    \phi(\vec{x}_i) &= G \sum_{j \neq i} m_j g(\vec{x}_i - \vec{x}_j) \\
    \vec{F}_i &= -\nabla_{x_i}\phi(\vec{x}_i) = -G \sum_{j \neq i} m_j \nabla_{x_i} g(\vec{x}_i - \vec{x}_j)
\end{align}

\subsection{The Fast Multipole Method}
The FMM is a hierarchical algorithm for efficiently evaluating interactions between particles \citep{greengard_rokhlin_1987}. Here we consider the Cartesian formulation of the FMM \citep{dehnen_2000,dehnen_2002}, generalized to arbitrary translation-invariant, spherically symmetric kernels. Consider two distinct sets of source points $S$ and destination points $D$. The core idea of the FMM is that interactions between points in $S$ and $D$ can be approximated by a two-sided expansion around the source reference centre $\vec{x}_S$ and the destination reference centre $\vec{x}_D$. The expansion coefficients can be used to summarize the effect of numerous interactions and greatly reduce the particle-level work that is required.

To simplify the combinatorial math, we use multi-index notation. For a tuple of non-negative integers $\vec{n} = (n_x, n_y, n_z)$, we define
\begin{align*}
    |\vec{n}| &= n_x + n_y + n_z \\
    \vec{n}! &= n_x! \, n_y! \, n_z! \\
    \vec{x}^{\vec{n}} &= x^{n_x} y^{n_y} z^{n_z} \\
    \begin{pmatrix} \vec{n} \\ \vec{k}  \end{pmatrix} &= \frac{\vec{n}!}{\vec{k}! (\vec{n} - \vec{k})!}
\end{align*}
Consider the Cartesian derivatives of our convolution kernel $g$
\begin{align}
    D_{\vec{n}} &= \frac{\partial^{|\vec{n}| }g(\vec{x})}{\partial \vec{x}^{\vec{n}}}
\end{align}
for example $D_{(2,0,1)} = \frac{\partial^3 g(\vec{x})}{\partial x_0^2 \partial x_2}$. This Cartesian derivative tensor $D$ is efficiently evaluated up to a given order $|\vec{n}| \leq p$ through the recurrence formula explained in \citet{tausch_2003}. For this kernel derivatives of the form
\begin{align}
    K_n(r) &:= \left(\frac{1}{r} \frac{\partial }{\partial r} \right)^n g(r)
\end{align}
are needed and \textsc{jz-fmm} permits extending the code for new convolution kernels by defining these derivatives in a simple \textsc{CUDA} function. The Plummer kernel can itself be evaluated very efficiently through the recursion
\begin{align}
    K_0(r) &= -\frac{1}{\sqrt{r^2 + \epsilon^2}} \\
    K_n(r) &=  -\frac{2n - 1}{r^2 + \epsilon^2} K_{n-1}(r)
\end{align}
and the normal gravitational kernel can use the same formula with $\epsilon = 0$.

The FMM approximates the kernel function around the location of a source node $\vec{x}_S$ and a destination node $\vec{x}_D$ through a two-sided expansion. Let $\vec{x}_i$ be a particle close to $\vec{x}_D$ and $\vec{x}_j$ one close to $\vec{x}_S$. Then we may approximate their interaction as
\begin{align}
    g(\vec{x}_j - \vec{x}_i) &\approx \sum_{|\vec{n}| \leq p} \frac{1}{\vec{n}!} ((\vec{x}_j - \vec{x}_S)  - (\vec{x}_i - \vec{x}_D))^{\vec{n}} D_{\vec{n}}(\vec{x}_S - \vec{x}_D)
\end{align}
where $D_{\vec{n}}$ = $D_{\vec{n}}(\vec{x}_S - \vec{x}_D)$ and $p$ denotes the highest derivative that is considered in the expansion. Using the Binomial theorem, the source and destination side can be separated
\begin{align}
    g(\vec{x}_j - \vec{x}_i) &\approx \sum_{|\vec{n}| \leq p} \frac{1}{\vec{n}!} D_{\vec{n}} \sum_{\vec{k} \leq \vec{n}} (-1)^{|\vec{k}|} \begin{pmatrix}
        \vec{n} \\ \vec{k}  \end{pmatrix} (\vec{x}_j - \vec{x}_S)^{\vec{n}-\vec{k}} (\vec{x}_i - \vec{x}_D)^{\vec{k}} \nonumber \\
        &= \sum_{\vec{k} \leq \vec{n} \leq p} \frac{(-1)^{|\vec{k}|}}{\vec{k}! (\vec{n} - \vec{k})!} (\vec{x}_i - \vec{x}_D)^{\vec{k}}  D_{\vec{n}} (\vec{x}_j - \vec{x}_S)^{\vec{n}-\vec{k}} \nonumber  \\
        &= \sum_{\vec{k} \leq p} (\vec{x}_i - \vec{x}_D)^{\vec{k}} \frac{(-1)^{|\vec{k}|}}{\vec{k}! } \sum_{\vec{n} \leq p - \vec{k}} \frac{1}{\vec{n}!}  D_{\vec{n} + \vec{k}} (\vec{x}_j - \vec{x}_S)^{\vec{n}} \label{eqn:fmm}
\end{align}
where in the last step we rearranged the summation space by offsetting $\vec{n} - \vec{k} \rightarrow \vec{n}$. Equation \eqref{eqn:fmm} forms the core of the FMM. It is useful to give some labels to the different operations. We may write the evaluation (for a sum of source particles) as a chain of operations
\begin{align}
    \phi(\vec{x}_i) &= G \sum_j m_j g(\vec{x}_j - \vec{x}_i) \\
                    &= L_{\vec{0}} \left(\vec{x}_i - \vec{x}_D, \mat{L}_D \left(\vec{x}_S - \vec{x}_D, \mat{\sum_j Q_S (\vec{x}_j, \mat{Q}_j)} \right) \right)
\end{align}
Reading the operations from inside to outside, we have a multipole to multipole (M2M) translation operator $\vec{Q}_S$, a multipole to local evaluation (M2L) operation $\vec{L}_D$ and a local to local (L2L) translation $\vec{L}_0$, that can be expressed in index notation as follows.
The multipole translation reads for monopole inputs
\begin{align}
    Q_{S,\vec{n}} &=  m_j (\vec{x}_j - \vec{x}_S)^{\vec{n}} \label{eqn:monopole2m}
\end{align}
or more generally for arbitrary multipole inputs
\begin{align}
    Q_{S,\vec{n}} (\vec{x}_j, \mat{Q}_{j}) &= \sum_{\vec{k} = \vec{0}}^{\vec{k} \leq \vec{n}} \begin{pmatrix} \vec{n} \\ \vec{k} \end{pmatrix} (\vec{x}_j - \vec{x}_S)^{\vec{n} - \vec{k}} Q_{j,\vec{k}} \label{eqn:m2m}
\end{align}
Next, we have the multipole to local operation
\begin{align}
    L_{D,\vec{k}}(\mat{Q}) &= G \frac{(-1)^{|\vec{k}|}}{\vec{k}!} \sum_{ |\vec{k} + \vec{n}| \leq p} \frac{1}{\vec{n}!}  D_{\vec{k} + \vec{n}} Q_{S,\vec{n}} \label{eqn:m2l}
\end{align}
with $D_{\vec{k} + \vec{n}}$ evaluated at $\vec{x} = \vec{x}_S - \vec{x}_D$ so that the expansion coefficients $L$ are evaluated at the destination node centre. Finally, we have the L2L operation that shifts the expansion coefficients from the destination centre $\vec{x}_D$ to a new location $\vec{x}_i$:
\begin{align}
    L_{\vec{n}} (\vec{x}_i, \mat{L}_{D}) &= \sum_{\vec{k} = \vec{n}}^{|\vec{k}| \leq p} \begin{pmatrix} \vec{k} \\ \vec{n} \end{pmatrix} (\vec{x}_i - \vec{x}_D)^{\vec{k} - \vec{n}} L_{D,\vec{k}} \label{eqn:l2l}
\end{align}
where 
\begin{align}
    L_{\vec{n}} (\vec{x}_i) &= \frac{1}{\vec{n}!} \frac{\partial^{|\vec{n}|} \phi}{\partial x_i^{n_x} \partial y_i^{n_y} \partial z_i^{n_z}}
\end{align}
so that the potential corresponds to $\vec{n} = 0$ and forces can be evaluated through components with $|\vec{n}| = 1$.

Two useful relations for spatial derivatives of expansion coefficients are
\begin{align}
    \frac{\partial Q_{S,i,\vec{n}}}{\partial x_{i,\alpha}}  &= n_\alpha Q_{S,i,\vec{n} - \vec{e}_\alpha} \label{eqn:m2m_deriv} \\
    \frac{\partial L_{\vec{n}}}{\partial x_{i,\alpha}}  &= (n_\alpha + 1) L_{\vec{n} + \vec{e}_\alpha} \label{eqn:loc_deriv}
\end{align}
where $\vec{e}_\alpha$ is the unit multi-index with the $\alpha$ component equal to 1 and for clarity $Q_{S,i}$ only denotes a single child's contribution to the multipole at $S$.

Conceptually the FMM with a tree proceeds as follows \citep[e.g.][]{greengard_rokhlin_1987,carrier_1988,dehnen_2002}:
\begin{itemize}
    \item A tree is constructed that groups particles into leaf-nodes at the finest level and hierarchically into coarser nodes at higher levels of the tree
    \item Leaf multipoles are calculated from particles as in equation~\eqref{eqn:monopole2m}.
    \item Coarser nodes' multipoles are calculated in an upward pass (from finer to coarser levels) by translating child node multipoles to their parents' centre and adding them together as in equation~\eqref{eqn:m2m}
    \item In a downward pass M2L translations are used to calculate node-node interactions via equation~\eqref{eqn:m2l}. Only well separated interactions are evaluated, whereas too close interactions are flagged for evaluation at a finer level as will be explained in Section~\ref{sec:dualtree}.
    \item Local expansions from coarser levels are advected to finer levels via the L2L operator from equation~\eqref{eqn:l2l}.
    \item Finally, the local expansions are translated from leaf centres to particle positions via equation~\eqref{eqn:l2l}. The potential and forces can be read out as the local expansion coefficients up to order 1.
    \item Interactions that cannot be evaluated as node-node interactions at the finest level are evaluated through grouped direct summation.
\end{itemize}

\subsection{Automatic differentiation}

Consider a chain of functions of the form
\begin{align}
    \text{Loss} = J(\vec{x}_n(\vec{x}_{n-1}(...\vec{x}_1(\vec{x}_0)...)))
\end{align}
where $J$ is a loss function with scalar output and $\vec{x}_n$ denotes a function that maps a previous input $\vec{x}_{n-1}$ to a new output. The most common goal of optimization techniques is to find a set of input parameters $\vec{x}_0$ that minimize such a loss function by iteratively moving the inputs in the directions of gradients $\partial J / \partial \vec{x}_0$.

Such gradients can be obtained through repeated evaluation of the chain formula
\begin{align}
    \frac{\partial J}{\partial \vec{x}_0} &= \frac{\partial{J}}{\partial \vec{x}_n} \frac{\partial{\vec{x}}_n}{\partial \vec{x}_{n-1}} ... \frac{\partial{\vec{x}}_1}{\partial \vec{x}_{0}} \label{eqn:backpropchain}
\end{align}
Note that derivatives of the form $\frac{\partial{\vec{x}}_n}{\partial \vec{x}_{n-1}}$ are full Jacobians whose explicit storage is generally quadratic in the number of degrees of freedom. The core idea of reverse-mode automatic differentiation, also called back-propagation or adjoint differentiation, is that these Jacobians never need to be evaluated explicitly; instead, gradients are propagated through vector-Jacobian product (VJP) operations for the intermediate steps \citep{linnainmaa_1976,baydin_2018}.

\begin{figure}
    \includegraphics[width=\columnwidth]{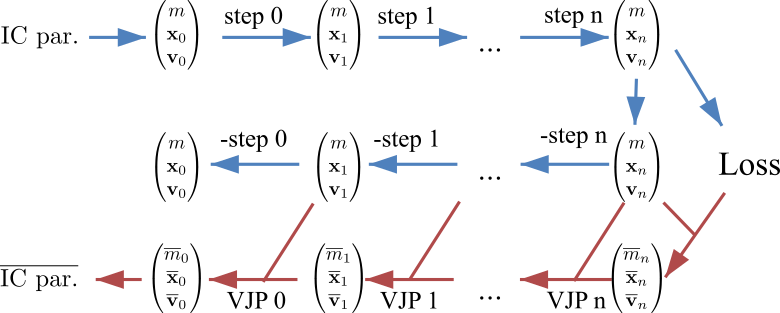}
    \caption{Illustration of how initial condition gradients are obtained through a differentiable N-body simulation.}
    \label{fig:nbodyvjp}
\end{figure}

Let us define adjoint variables or 'sensitivities'
\begin{align}
    \bar{\vec{x}}_n := \frac{\partial J}{\partial \vec{x}_n}
\end{align}
as the derivative of the loss function with respect to some intermediate state $\vec{x}_n$. We can evaluate equation \eqref{eqn:backpropchain} step by step going from left to right (i.e. backwards) and applying backwards advection steps of the form
\begin{align}
    \bar{\vec{x}}_{n-1} = \bar{\vec{x}}_{n} \frac{\partial{\vec{x}}_n}{\partial \vec{x}_{n-1}}
\end{align}
Such VJPs can be evaluated for many types of operations without actually instantiating the Jacobian. Standard usage of \textsc{jax} allows defining a set of functions in forward evaluation as a graph that can automatically be transformed to a chain of VJP rules when gradients are requested \citep{jax_2018}. However, to achieve optimal performance, we are using a \textsc{CUDA} extension through \textsc{jax}'s foreign function interface (FFI) to evaluate the FMM. Such extensions cannot be automatically differentiated by \textsc{jax} and need custom derivative rules. We will show here that defining such a custom VJP rule for the FMM is particularly clear and simple.

Figure~\ref{fig:nbodyvjp} illustrates how the gradient calculation for an entire N-body simulation operates in practice. First the simulation is run with forward time-stepping towards the final output distribution in which a loss function is defined (e.g. to compare with observations). Gradients of the loss with respect to the particle masses, positions and velocities are inferred at the final time and then backwards advected with the VJP rule of a time step up to the initial conditions. This VJP requires knowledge of the position and velocity state at the given time which can either be saved from the forward integration (at a substantial memory cost) or -- as done here -- reconstructed through a simultaneous backward integration of the reversible N-body system \citep{li_2024,List_2026}. For this it is desirable to have a bit-perfect backwards integration, which can be achieved with a reproducible force calculation plus an integer lattice integrator approach \citep{miller_1970,syer_tremaine_1995,mocz_succi_2017,rein_tamayo_2018}.

The VJP rule of an N-body time-step (e.g. a DKD step) is quite simple \citep[see e.g.][]{li_2024,List_2026} so we will only discuss the most difficult component here, which is the VJP of the force calculation. Beyond that, we discuss in Appendix~\ref{app:nbodyjvpvjp} an interesting symmetry between the VJP and the Jacobian vector product (JVP) of the forces and of the symplectic N-body system overall. It implies that a JVP through the forward integration can be evaluated as a VJP through the inverse integration after appropriately rotating the input and output phase-space coordinates, allowing forward derivatives to be evaluated with codes that only implement reverse derivatives, and vice versa.

\subsection{VJP of the Fast Multipole Method}


To define the VJP rule for the FMM, we need to evaluate how sensitivities of the output local expansions $\vec{\bar{L}}$ (including potential, forces and possibly higher order terms) translate to sensitivities of the input positions $\vec{x}$, masses $\vec{m}$.

Here, we show that the core component of the VJP of the FMM may be implemented as a standard FMM with modified inputs where multipoles of the input sources get replaced by the local expansions of the adjoint output variables. 

We assume that the tree structure and node positions are static during differentiation. This gives the correct gradients for a geometrically centred tree (apart from discontinuities as discussed below), but only approximate gradients for a mass-centred tree (where node centres should have a derivative with respect to particle positions). While mass-centring is supported in \textsc{jz-fmm}, the default mode is geometric centring, since we have not observed any notable accuracy benefits of mass-centring for $p>1$. It is worth noting that the force-field defined by the FMM is naturally discontinuous at node boundaries and derivatives are ill-defined at those node boundaries. However, we will show in Section~\ref{sec:simgrad} that this does not compromise the convergence of simulation gradients.





\subsubsection{Adjoint of L2L maps to M2M}

The adjoint operation of the L2L operator in Equation~\eqref{eqn:l2l} needs to calculate an output sensitivity $\vec{\bar{L}}_D$ for the destination node, and an additional sensitivity for the translation location $\vec{\bar{x}}_i$ if that variable is considered differentiable. This means for node-node translations we only need to consider $\vec{\bar{L}}_D$, whereas for node-particle translations we also need to consider $\vec{\bar{x}}_i$.

The sensitivity for particle positions reads
\begin{align}
    \bar{x}_{v, \alpha} &= \sum_i \sum_{\vec{n}} \bar{L}_{i,\vec{n}} \frac{\partial L_{i, \vec{n}}}{\partial x_{v,\alpha}} \\
    &= \sum_{\vec{n} + \vec{e}_\alpha \leq p} (n_\alpha + 1) \bar{L}_{v,\vec{n}} L_{v, \vec{n} + \vec{e}_\alpha}
    \label{eqn:explicitvjpl2l}
\end{align}
which follows directly from equation~\eqref{eqn:loc_deriv}. This term describes how the potential and force derivatives of a particle are given by the force and tidal field at the particle's location (when all other particles are considered fixed).
To get the sensitivity of the input local expansion of the L2L translation, we accumulate the sensitivity from the child ($\bar{L}_i$) into the parent ($\bar{L}_D$). By the chain rule
\begin{align}
    \bar{L}_{D, \vec{k}} &= \sum_i \sum_{\vec{n}} \bar{L}_{i, \vec{n}} \frac{\partial L_{i, \vec{n}}}{\partial L_{D, \vec{k}}}
\end{align}
the partial derivative is:
\begin{align}
    \frac{\partial L_{i, \vec{n}}}{\partial L_{D, \vec{k}}} &= 
    \begin{cases} 
        \binom{\vec{k}}{\vec{n}} (\vec{x}_i - \vec{x}_D)^{\vec{k} - \vec{n}} & \text{if } \vec{k} \ge \vec{n} \\
        0 & \text{otherwise}
    \end{cases}
\end{align}
Substituting this back, we sum over valid $\vec{n} \leq \vec{k}$:
\begin{align}
    \bar{L}_{D, \vec{k}} &= \sum_i \sum_{\vec{n}=\vec{0}}^{\vec{n} \leq \vec{k}} \binom{\vec{k}}{\vec{n}} (\vec{x}_i - \vec{x}_D)^{\vec{k} - \vec{n}} \bar{L}_{i, \vec{n}}
\end{align}
Comparing this to Eq. (\ref{eqn:m2m}), we see that $\bar{L}$ transforms exactly like a multipole moment $\mat{Q}$. Thus, this second part of the adjoint of the L2L operation is mathematically identical to an M2M translation of the adjoint variables.

\subsubsection{Adjoint M2L (Interaction)}
In the forward pass, the M2L operator (Eq.~\eqref{eqn:m2l}) converts a multipole expansion $\mat{Q}_S$ at a source node centre $\vec{x}_S$ to a local expansion $\mat{L}_D$ at a destination centre $\vec{x}_D$.
\begin{align}
    L_{D, \vec{k}} &= G \frac{(-1)^{|\vec{k}|}}{\vec{k}!} \sum_{\vec{n}} \frac{1}{\vec{n}!} D_{\vec{k} + \vec{n}}(\vec{x}_S - \vec{x}_D) Q_{S, \vec{n}}
\end{align}
The adjoint operation computes the influence of the destination adjoints $\bar{L}_D$ on the source multipole adjoints $\bar{Q}_S$:
\begin{align}
    \bar{Q}_{S, \vec{n}} &= \sum_{\vec{k}} \bar{L}_{D, \vec{k}} \frac{\partial L_{D, \vec{k}}}{\partial Q_{S, \vec{n}}}
\end{align}
Computing the partial derivative:
\begin{align}
    \frac{\partial L_{D, \vec{k}}}{\partial Q_{S, \vec{n}}} &= G \frac{(-1)^{|\vec{k}|}}{\vec{k}! \vec{n}!} D_{\vec{k} + \vec{n}}(\vec{x}_S - \vec{x}_D)
\end{align}
Substituting this into the sum:
\begin{align}
    \bar{Q}_{S, \vec{n}} &= \frac{G}{\vec{n}!} \sum_{\vec{k}} \frac{(-1)^{|\vec{k}|}}{\vec{k}!} D_{\vec{k} + \vec{n}}(\vec{x}_S - \vec{x}_D) \bar{L}_{D, \vec{k}}
\end{align}
Using the symmetry property of the kernel derivatives $D_{\vec{p}}(\vec{r}) = (-1)^{|\vec{p}|} D_{\vec{p}}(-\vec{r})$, we have:
\begin{align}
    D_{\vec{k}+\vec{n}}(\vec{x}_S - \vec{x}_D) &= (-1)^{|\vec{k}|+|\vec{n}|} D_{\vec{k}+\vec{n}}(\vec{x}_D - \vec{x}_S)
\end{align}
Substituting this back yields:
\begin{align}
    \bar{Q}_{S, \vec{n}} &= G \frac{(-1)^{|\vec{n}|}}{\vec{n}!} \sum_{\vec{k}} \frac{1}{\vec{k}!} D_{\vec{k} + \vec{n}}(\vec{x}_D - \vec{x}_S) \bar{L}_{D, \vec{k}}
\end{align}
This equation has the exact same structure as the forward M2L (Eq. \ref{eqn:m2l}), but with roles swapped ($\vec{\bar{L}}_D$ acts as source multipole, $\vec{\bar{Q}}$ acts like the destination expansion).

\subsubsection{Adjoint M2M (Downward Pass)}
In the forward pass, the M2M operator (Eq. \ref{eqn:m2m}) shifts a multipole expansion $\mat{Q}_j$ from one expansion centre $\vec{x}_j$ to another one $\vec{x}_S$.
\begin{align}
    Q_{S, \vec{n}} &= \sum_{\vec{k}=\vec{0}}^{\vec{n}} \binom{\vec{n}}{\vec{k}} (\vec{x}_j - \vec{x}_S)^{\vec{n} - \vec{k}} Q_{j, \vec{k}}
\end{align}
For the adjoint, we may have two types of terms:
(1) The derivative with respect to the input multipole values. 
(2) The explicit derivative with respect to the position of the input multipoles $\vec{x}_j$.

The first term is given by 
\begin{align}
    \bar{Q}_{j, \vec{k}} &= \sum_{\vec{n}} \bar{Q}_{S, \vec{n}} \frac{\partial Q_{S, \vec{n}}}{\partial Q_{j, \vec{k}}}
\end{align}
The partial derivative selects terms where $\vec{n} \ge \vec{k}$:
\begin{align}
    \frac{\partial Q_{S, \vec{n}}}{\partial Q_{j, \vec{k}}} &= \binom{\vec{n}}{\vec{k}} (\vec{x}_j - \vec{x}_S)^{\vec{n} - \vec{k}}
\end{align}
Substituting back:
\begin{align}
    \bar{Q}_{j, \vec{k}} &= \sum_{\vec{n}=\vec{k}}^{p} \binom{\vec{n}}{\vec{k}} (\vec{x}_j - \vec{x}_S)^{\vec{n} - \vec{k}} \bar{Q}_{S, \vec{n}}
\end{align}

Comparing this to Eq. (\ref{eqn:l2l}), we see that the indices and binomial structure exactly match the Local-to-Local translation. Thus, the adjoint of the M2M operation is mathematically identical to an L2L translation of the adjoint variables.

Term (2) again only appears for particle-node translations, since node positions are considered static. For this explicit sensitivity we find
\begin{align}
    \bar{x}_{v, \alpha} &= \sum_{\vec{n}} \bar{Q}_{S,\vec{n}} \frac{\partial Q_{S,v,\vec{n}}}{\partial x_{v,\alpha}} \nonumber \\
    &= \sum_{\vec{n}  \leq p} n_\alpha \bar{Q}_{S,v,\vec{n}} Q_{S,v, \vec{n} - \vec{e}_\alpha}  \nonumber \\
    &= \sum_{\vec{n} + \vec{e}_\alpha \leq p} (n_\alpha + 1) \bar{Q}_{S,\vec{n} + \vec{e}_\alpha} Q_{S,v, \vec{n}}
    \label{eqn:explicitvjpm2m}
\end{align}
where in the last step we shifted the indices, to clarify that this is structurally identical to equation~\eqref{eqn:explicitvjpl2l}, but with switched roles of the sensitivity and original moments. Since the adjoint multipole translation behaves exactly like an L2L operation, we can simply get the multipole sensitivity directly at the particle location first and then evaluate~\eqref{eqn:explicitvjpm2m} with $Q_{S,v, \vec{0}} = m_v$ for monopoles.

We summarize that the VJP of the FMM can easily be evaluated in the following steps:
\begin{itemize}
    \item The local expansion at particle locations is kept from the normal FMM evaluation up to one order higher than the normal output would require. The effect that each particle's position has on its own potential and force is evaluated as in equation~\eqref{eqn:explicitvjpl2l}. This gives a first contribution to the position sensitivity.
    \item We evaluate the full FMM, but with the tangents of the local expansion terms at particle locations as inputs (rather than mass moments). For example, potential sensitivities act as monopoles and force sensitivities as dipoles. The output will be mass multipole sensitivities at particle locations (instead of local expansions).
    \item For monopoles the mass sensitivities are given by the zeroth component whereas position sensitivities can be evaluated with equation~\eqref{eqn:explicitvjpm2m}.
\end{itemize}

\section{\textsc{jz-fmm} implementation}

We have designed \textsc{jz-fmm} from scratch with a pure GPU computation perspective in mind. This is motivated by the high performance that is needed for solving reconstruction problems that may require thousands of simulation gradient evaluations -- something that is much more feasible with the superior computational throughput of GPUs (with respect to CPUs). When designing the code, we have paid particular attention to the strengths and weaknesses of GPU hardware. Most importantly this means
\begin{itemize}
    \item Minimizing device memory access / maximizing the computation that can be done per memory access.
    \item Ensuring thread groups read and reuse memory collaboratively.
    \item Keeping device memory access coalesced wherever possible.
    \item Keeping frequently updated variables in registers.
    \item Avoiding branches in the control flow between threads.
    \item Keeping the number of kernel launches low.
    \item Avoiding atomic addition operations for launch-independent bit-perfect reproducibility.
\end{itemize}
The required key-concepts are the bottom-up tree structure (briefly described in Section~\ref{sec:zorder} and in more detail in \citet{stuecker_2026}) and the non-recursive formulation of the dual tree walk described in Section~\ref{sec:dualtree}. Further, we will describe our implementation of multi-device parallelization in Section~\ref{sec:multidevice} and give an overview of more detailed implementation notes in Section~\ref{sec:implnotes}.

\subsection{Z-order tree}\label{sec:zorder}
The tree structure in \textsc{jz-fmm} is adopted from \textsc{jz-tree} \citep{stuecker_2026}. We only briefly mention the core concepts here, but refer to the corresponding article for an in-depth explanation.

\begin{figure*}
    \centering
    \begin{minipage}[c]{0.47\textwidth}
        \centering
        \includegraphics[width=\linewidth]{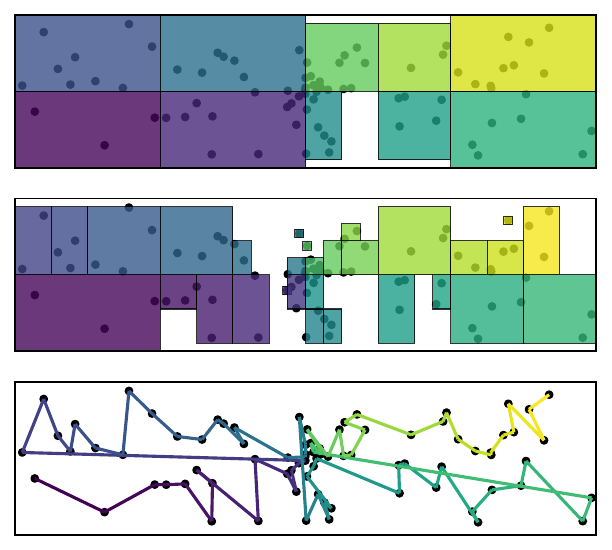}
    \end{minipage}\hfill
    \begin{minipage}[c]{0.49\textwidth}
        \centering
        \includegraphics[width=\linewidth]{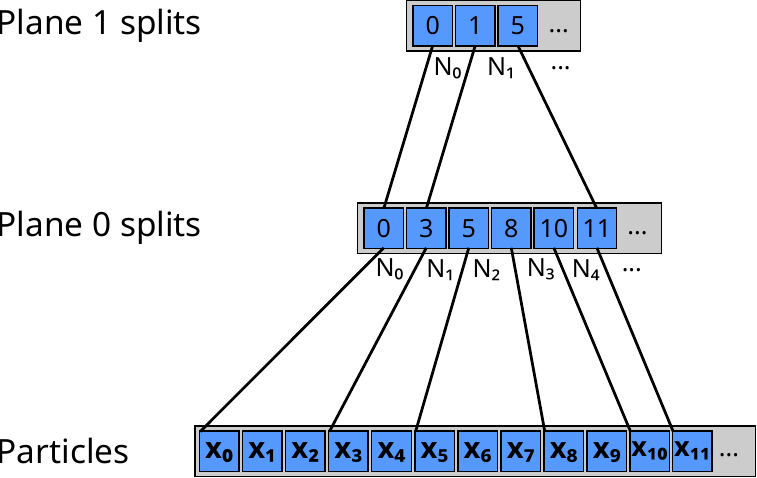}
    \end{minipage}
    \caption{Illustration of the plane-based tree hierarchy (left) and its data layout (right).
    The particle distribution is sorted in z-order (bottom) and to define leaf (plane 0) nodes, splitting points are found to assure that each leaf has $\leq \mathrm{max\_leaf\_size}$ particles (here 4). Coarser planes are defined based on the previous planes' splitting points with successively larger particle count limits (here 12 at plane 1). Note that some nodes extend beyond the displayed range. Importantly for GPU performance, children of a given node are always contiguous in memory.
    }
    \label{fig:treeplanes}
\end{figure*}

The core idea is to sort a set of particles along a space-filling curve and then to select splitting points between particles to define successively coarser nodes on the particle distribution. The tree is therefore built in a bottom-up manner, exposing substantially more fine-grained parallelism during construction and making the approach well suited to GPUs \citep{karras_2012,bedorf_2012}. This contrasts with traditional gravitational tree codes which construct their tree by recursively splitting nodes in a top-down approach \citep{barnes_hut_1986, dikaiakos_stadel_1996, springel_2005,potter_2017, springel_2021}.

Here, we choose to sort particles along the z-order (or Morton) curve \citep{morton_1966}. This has most of the same benefits as the more commonly used Peano Hilbert curve \citep[see e.g.][]{springel_2005}, but it is notably simpler and can more easily be generalized to infinite domains. Two particles $\vec{x}_A$ and $\vec{x}_B$ are in z-order if their (hypothetical) bit-interleaved representations are in order. Particles are sorted in z-order with a merge sort that uses a direct comparison operation on the full position vectors.  This allows the sort to perfectly respect any coordinate differences that can be represented in floating point precision, and it is therefore unnecessary to define a finite domain in \textsc{jz-fmm}.

We define the smallest z-tree node that contains two position vectors $\vec{x}_A$ and $\vec{x}_B$ as the spatial region where all vectors share the same leading bits as $\vec{x}_A$ and $\vec{x}_B$ in a (hypothetical) bit-interleaved representation. A set of nodes that covers all particles may be represented through a set of splitting points
\begin{align}
    \vec{\textbf{spl}} = (\text{spl}[0], \text{spl}[1], ...)
\end{align}
so that each node $i$ contains the particles with indices in range $\text{spl}[i] ... \text{spl}[i+1]-1$. Note that the splitting points cannot be chosen arbitrarily, but they must guarantee that nodes do not overlap. We refer to a segmentation of all particles into nodes so that each particle is part of exactly one node as a \emph{tree-plane}.

We can define a coarser tree-plane by choosing a sub-set of splitting points from a finer tree-plane. For example, given a leaf-plane $\vec{\textbf{spl}}^{(0)}$, a coarser plane $\vec{\textbf{spl}}^{(1)}$ can be defined so that level 1 node $i$ contains the level 0 nodes in range $\text{spl}^{(1)}[i] ... \text{spl}^{(1)}[i+1]-1$, corresponding to particles in range $\text{spl}^{(0)}[\text{spl}^{(1)}[i]] ... \text{spl}^{(0)}[\text{spl}^{(1)}[i+1]]-1$. This is illustrated in the right panel of Figure~\ref{fig:treeplanes}.

In \textsc{jz-fmm} and \textsc{jz-tree} we define the tree planes as the z-order segmentations with the largest possible nodes so that the number of particles in a level $p$ node is at most
\begin{align}
    n \leq \text{max\_leaf\_size} \cdot c^p
\end{align}
where $\text{max\_leaf\_size}$ (default 32) and $c$ (default 6) are parameters of the code. In the left panel of Figure~\ref{fig:treeplanes} we illustrate the first two planes that are obtained for a 2D particle distribution with $\text{max\_leaf\_size}=4$ and $c=3$. Typically, we only coarsen tree planes until the expected number of nodes is less than $1024$ (another parameter in the code), so that the top level of the tree already allows work to be efficiently parallelized on GPUs. Typically this leads to a quite small number of planes -- something that allows the FMM to be evaluated in a very small number of kernel dispatches.

The plane-based tree hierarchy has a few notable differences compared to more commonly used oct-tree or kd-tree layouts:
\begin{itemize}
    \item The tree has the same depth everywhere.
    \item Nodes may have a variable number of unequally sized children.
    \item Nodes on the same tree level may have different sizes.
    \item Some nodes may only contain a single particle (and have effectively 0 extent).
    \item A node may have itself as the only child on the next finer level.
    \item All particles are part of a node, but empty space may or may not be part of a node.
    \item Nodes may have axis ratios of 2:1.
\end{itemize}
The benefits of this flexible structure are borne out in practice (see Section~\ref{sec:performance}). Further, we note that the plane-based tree can be built extremely efficiently on GPU. The dominant cost is the sort, which is a highly performant operation on GPU.

After a tree is constructed in \textsc{jz-fmm}, multipoles are determined for every node. On the leaf level this follows equation~\eqref{eqn:monopole2m} and then for each higher level tree-plane child multipoles are shifted to their parents' centres following equation~\eqref{eqn:m2m}. This requires one kernel dispatch per tree-plane. Multipoles are stored per plane in dense arrays that follow the layout of the splitting points, but have one fewer element each. By default, $\textsc{jz-fmm}$ uses the geometric centres of nodes as expansion points, but it is also possible to use a mass-centred approach (requiring one additional upwards pass).

One noteworthy implementation detail is that to remain compatible with \textsc{jax}'s just-in-time (jit) compilation, it is necessary that no allocations are data-dependent. However, the required number of nodes \emph{is} data-dependent. We follow the \textsc{jz-tree} approach to predict moderately generous allocation sizes at jit-compile time and to dynamically keep track of the actually filled data sizes at run time to mask out invalid elements.

\subsection{Dual tree walk}\label{sec:dualtree}

The FMM dual tree walk in \textsc{jz-fmm} mirrors the strategy presented in \textsc{jz-tree} for k nearest neighbour search and friends-of-friends clustering \citep[see also][]{dehnen_2002}. In short: Interactions are evaluated plane by plane from the coarsest to the finest level while keeping track of nodes' local expansions and a list of interactions that need to be evaluated on the next finer level. Interactions that cannot be evaluated at the finest level are evaluated through direct summation in a final step.

Interactions are stored as node indices in a dense array $\textbf{ilist}^{(p)}$ plus a set of splitting points $\textbf{ispl}^{(p)}$ that are defined so that node $i$ on plane $p$ needs to receive interactions from all nodes in the interaction list at indices $\text{ispl}^{(p)}[i] ... \text{ispl}^{(p)}[i+1]-1$. An entry means that the corresponding interaction is meant to be opened, i.e. all children of node $i$ need to interact with all children of all the nodes in its list. As described in \citet{stuecker_2026}, we initialize a dense interaction list (every node interacting with every other node) at one level higher than the top-level of the tree $p+1$ and zero out all expansion coefficients at that level. 

The level $p$ evaluation receives the level $p+1$ interaction list and expansion coefficients as an input and outputs a level $p$ interaction list and expansion coefficients. The higher level expansion coefficients are simply used for the L2L translation from equation~\eqref{eqn:l2l} to the children at level $p$. Further, we define an M2L kernel that either evaluates interactions via equation~\eqref{eqn:m2l} or inserts them into the level $p$ interaction list, depending on whether an opening criterion is fulfilled. In \textsc{jz-fmm} we use as a default opening criterion
\begin{align}
    \text{open if:} \Leftrightarrow \lvert \vec{l}_A + \vec{l}_B \rvert \geq \theta_{\mathrm{max}} \lvert \vec{x}_A - \vec{x}_B \rvert
\end{align}
where $\vec{l}_A$ and $\vec{l}_B$ are the nodes' half-extent vectors, $\vec{x}_A$ and $\vec{x}_B$ are the node centres and $\theta_{\mathrm{max}}$ is the largest permitted opening angle. For geometrically centred nodes the FMM is convergent with expansion order for $\theta_{\mathrm{max}} < 1$ \citep{engblom_2011}, with the criterion ensuring that accepted node pairs do not overlap. The default choice is $\theta_{\mathrm{max}} = 0.8$ with a high expansion order $p=5$. Note that our opening criterion is slightly different from other codes, due to the necessity to deal with unequally sized and non-cubic nodes. For cubic nodes with equal size it is slightly less conservative by a factor $\sqrt{3/4}$ than the one employed in \textsc{gadget4}~\citep{springel_2021} at a given $\theta_{\mathrm{max}}$.

\begin{figure}
\centering
\fbox{%
\begin{minipage}{0.96\columnwidth}
\small
\raggedright
\textbf{For one receiver parent node $A$:}
\begin{enumerate}
    \item One thread block is assigned to the parent node $A$.
    \item \textbf{Set up receiver children.} The $n_A$ children of $A$ have their centres and extents loaded into shared memory. The block threads are divided evenly over these children, assigning approximately $N_{\mathrm{threads}}/n_A$ threads to each child.
    \item \textbf{For each contiguous source chunk:} Every thread reads one source child $b$ from the interaction list of $A$, loading the block's source data collaboratively into shared memory.
    \begin{enumerate}
        \item \textbf{Classify interactions.} For every receiver child $a$, evaluate the opening criterion for each streamed source child $b$,
        \[
            \mathrm{Open}(a,b) \iff \lvert \vec{l}_a + \vec{l}_b \rvert \geq \theta_{\mathrm{max}} \lvert \vec{x}_a - \vec{x}_b \rvert .
        \]
        The threads collectively count the opened interactions and flag the source children that can instead be evaluated by M2L.
        \item \textbf{Evaluate accepted interactions.} For every receiver child $a$, the threads assigned to $a$ divide its flagged source children between them and accumulate independent partial contributions,
        \[
            L_a^{(p)} \mathrel{+}= \mathrm{M2L}(\vec{x}_b - \vec{x}_a, Q_b) .
        \]
    \end{enumerate}
    \item \textbf{After all source chunks:} The partial M2L contributions for each receiver child $a$ are summed and written to $L_a^{(p)}$. The opened-interaction counts are written out for the separate insertion kernel that builds the next-level interaction list.
\end{enumerate}
\end{minipage}%
}
\caption{Simplified outline of the M2L kernel. Threads collaborate on the child nodes of an opened parent node greatly reducing global memory access.}
\label{fig:m2l_kernel}
\end{figure}

Since the M2L kernel is (besides the direct summation kernel) the most performance critical component of an FMM implementation, it is worth highlighting in detail how to achieve a GPU friendly evaluation structure. Several aspects are of key importance here
\begin{itemize}
    \item GPUs execute threads in groups. Branches and recursion should be avoided to guarantee that all threads in one group execute the same instruction at a time.
    \item Memory access speed is often the most limiting factor. Reading data collaboratively (into shared memory) and reusing it across threads can greatly improve performance.
    \item Memory access is significantly faster if data is read in contiguous chunks across threads (memory coalescence).
\end{itemize}
We outline the structure of the M2L kernel in Fig.~\ref{fig:m2l_kernel}. The core idea is that all children of a given receiving parent node need to iterate over the same source nodes and can therefore read data collaboratively, greatly reducing the number of memory accesses. Further, children of a given source node are guaranteed to be contiguous in memory and source nodes are encountered in the interaction list in ascending order and will often form even larger contiguous segments, leading to great memory coalescence. Finally, we note that all interactions of a node are fully evaluated in one thread group, thus not requiring any atomic addition operations. This allows for bit-perfect reproducible results, independent of execution order and multiprocessor count, a feature that is quite important for making simulations reversible and hardware independent.

Finally, once the downward pass has reached the leaf plane, local expansions are read out at particle positions and the remaining interaction list is evaluated in a dedicated direct summation kernel. This kernel follows a similar structure as outlined in Figure~\ref{fig:m2l_kernel}, but with steps (a) and (b) simply replaced by a direct summation over the source particles. Again, we emphasize that per opened leaf, the particles in other interacting leaves only need to be read once (rather than once per particle).

\subsection{Multi-device parallelization}\label{sec:multidevice}

The parallelization in \textsc{jz-fmm} follows the pattern in \textsc{jz-tree}: After a global z-order sort with a sampling based initial domain split, domain boundaries are slightly adjusted to guarantee that domain splits only happen at top-level node boundaries. Building the tree then becomes a purely local problem and each device has a unique set of nodes. For the evaluation, we always keep receiving nodes local and only communicate source nodes (or source particles) where necessary. The interaction list additionally keeps track of the origin device of each (unique) appearing node and is initialized at the highest level for the local receiving nodes on each device with source indices over all global top-level nodes. Before each M2L kernel launch, the interaction list information is then used to request the source multipoles required for the evaluation. Similarly, for the direct summation kernel all child particles of the interacting leaves are requested in advance.

This communication pattern has a few notable properties:
\begin{itemize}
    \item Only a small number of global communication steps is required (a few per plane).
    \item The single-device and multi-device implementations are mostly identical, only differing in the way source nodes are obtained.
    \item The interaction list approach allows us to guarantee that every remote source node is only requested once.
    \item Storing all temporary requested source node/particle data needed for evaluation requires a notable amount of additional memory -- typically $O(1.1 - 1.2)$ times the size of the local multipole/particle data. Since this allocation factor needs to be predicted at jit-compile time, \textsc{jz-fmm} exposes a config argument for this that defaults to a notable margin (1.5).
    \item The evaluation order for a receiving node is perfectly independent of the number of devices used, thus achieving bit-perfect reproducibility even when different device counts are used.
\end{itemize}
The main drawback is the requirement of the rather large additional source node allocation to guarantee that all interactions can be evaluated in one pass. As highlighted in the beginning, our main optimization target is performance rather than memory, but future updates of \textsc{jz-fmm} may also include alternative, more memory-optimized evaluation patterns.

\subsection{Other implementation notes}\label{sec:implnotes}
The \textsc{jz-fmm} implementation is formulated in a rather general way, supporting a notable number of features:
\begin{itemize}
    \item Variable problem dimensionality (only dimensions 2 and 3 have been tested and are included in \textsc{PyPI} binaries.)
    \item Arbitrary multipole order (only $p=1-7$ have been tested and are included by default.)
    \item Float and double precision are supported.
    \item Since the direct summation kernel may easily accumulate a notable summation error, we provide the option to enhance its precision through Kahan summation \citep{Kahan1965}.
    \item We provide a simple pattern to allow replacing the convolution kernel -- which however, requires recompiling the \textsc{CUDA} kernels. By default, Plummer, 2D Plummer-equivalent and a softened distance kernel (see Section~\ref{sec:optimization}) are supported. The code could easily be adapted for use in other domains that require convolutions, e.g. radial basis function interpolation and kernel density estimates.
    \item The node-node part of the FMM allows evaluating local expansions at any order. However, the direct summation kernel is so far specialized to only compute potentials and forces, but it could easily be extended to higher derivatives such as tidal fields as well -- for example to support methods like the Geodesic Deviation Equation \citep{Vogelsberger_2011,Stuecker_2022}.
    \item The code follows a perfectly functional design. That means zero global variables are used, all relevant input parameters appear in the signature and outputs are return values.
    \item Every function is modular and could be used in a different context.
    \item The code is fully compatible with \textsc{jax.jit} and \textsc{jax.grad} to allow compilation into a fully optimized graph at a higher, user-controlled level.
    \item The code is compatible with user defined 'shard-maps'. For example, it would be possible to evaluate the FMM on a subset of devices while performing other types of computation on other devices.
    \item The code library includes a large number of unit tests and benchmarks of individual components. This allows users to easily verify the consistency and screen the performance impact of modifications to the code.
\end{itemize}
The flexibility in dimension, multipole order and data types is implemented through template parameters to guarantee that each kernel is properly optimized for each scenario. We have taken particular care to guarantee that all array like variables (e.g. multipoles or local expansions) can be represented in registers by avoiding any dynamic indexing through static loop-unrolling. In principle the code also supports compilation for higher order expansions $p \geq 8$, but for such cases the register space in the GPUs' multiprocessors turns out to be too small to hold all the required terms, leading the compiler to place arrays in device memory instead (also known as ``register spilling'') which may slow down such kernels by orders of magnitude. We will therefore not consider $p \geq 8$ here. We note that a (less memory-intense) spherical harmonics expansion may reduce such problems by using the trace-less nature of the gravitational kernel \citep{dehnen_2014}, but we did not attempt such an optimization here, to keep consistency with general kernel functions.

Finally, we note that in \textsc{jz-fmm} we have implemented the necessary features that allow running simple simulations
\begin{itemize}
    \item Standard Kick-Drift-Kick (KDK) integrator.
    \item Standard DKD integrator.
    \item Standard DKD integrator with integer lattice internal state \citep{miller_1970,syer_tremaine_1995,mocz_succi_2017,rein_tamayo_2018}. This method represents positions and velocities through an integer lattice in the time integration, but converts to floating point numbers for the FMM force evaluation. This allows perfectly reversible simulations, which is relevant for gradient computation.
    \item Differentiability for both DKD integrators.
    \item A flexible method for defining analytical external potentials (and automatic differentiation thereof).
\end{itemize}
With this \textsc{jz-fmm} can easily be used out-of-the-box for simple idealized N-body simulations. That said, we primarily intend this library to be used as a building block in larger differentiable N-body codes and therefore the focus here is on the FMM itself.

\section{Accuracy and performance}~\label{sec:performance}
We evaluate the accuracy and performance of \textsc{jz-fmm} for a variety of simple test problems. For accuracy checks we always compare to a direct summation evaluation of the same problem with the same softening kernel, but with Kahan summation enabled to minimize the accumulation error. All tests use 32-bit floating point precision, unless explicitly mentioned. All differentiable simulations use DKD integration, with the integer lattice variant used by default; all other simulations in \textsc{jz-fmm} use standard floating-point DKD.

All performance tests are run on booster nodes of the Leonardo cluster at CINECA \citep{turisini_2024_leonardo}, that have per node four NVIDIA A100-64 GPUs, 200 Gbps NVIDIA Mellanox high data rate (HDR) InfiniBand connection and one 32 core Intel Xeon Platinum 8358 processor.

\subsection{Convergence tests}
\begin{figure}
    \includegraphics[width=\columnwidth]{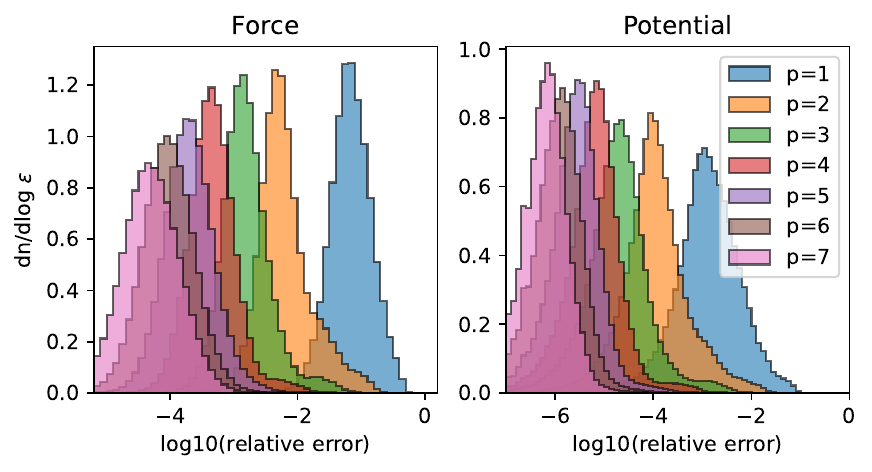}
    \caption{Relative force and potential errors of the FMM evaluated for a three-dimensional normal distribution of particles. The method is convergent with expansion order, yielding at $p=5$ relative force errors $\lesssim 10^{-3}$ and much smaller potential errors $\lesssim 10^{-5}$.}
    \label{fig:forceerrors}
\end{figure}

We sample $N=1024^2$ particles from an isotropic three-dimensional normal distribution. We calculate FMM forces and potentials with a Plummer softening with $\epsilon = 10^{-2} \sigma$ and estimate the relative errors per particle compared to the direct summation case as
\begin{align}
    \epsilon_F = \frac{\lVert \vec{F} - \vec{F}_{\mathrm{ref}} \rVert}{\lVert \vec{F}_{\mathrm{ref}} \rVert} \\
    \epsilon_\phi = \frac{\lvert \phi - \phi_{\mathrm{ref}} \rvert}{\lvert \phi_{\mathrm{ref}} \rvert}
\end{align}
We use the default parameters of the code (in particular $\theta_{\mathrm{max}} = 0.8$), but vary the expansion order between $p=1$ and $p=5$. The resulting error distributions are shown in Figure~\ref{fig:forceerrors}. Clearly, the method is convergent with expansion order, yielding at $p=5$ small relative force errors ($\epsilon_F \lesssim 10^{-3}$) and much smaller relative potential errors ($\epsilon_\phi \lesssim 10^{-5}$). We have also tried this test with a mass-centred expansion approach, but did not observe relevant improvements, thus the default choice of a geometrically centred tree.

\begin{figure}
    \includegraphics[width=\columnwidth]{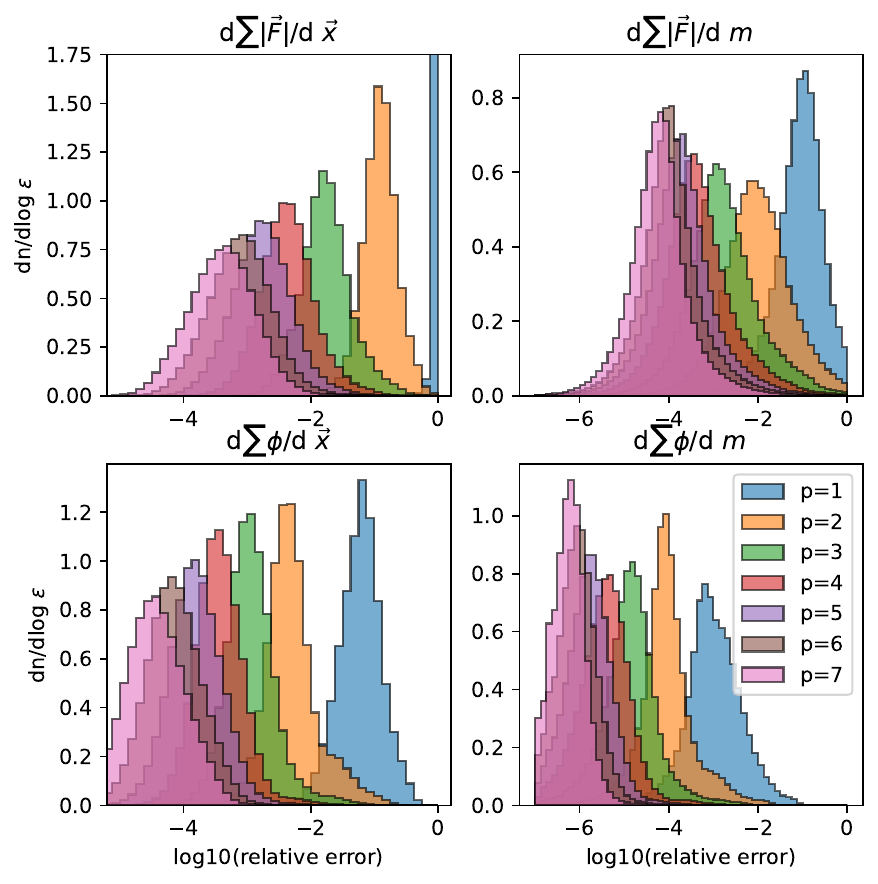}
    \caption{Relative differences of FMM gradients for a three-dimensional normal distribution of particles compared to direct summation. Gradients converge consistently, but at a lower convergence order depending on the type of the gradient.}
    \label{fig:graderrors}
\end{figure}

We evaluate the accuracy of the VJP for the same setup. Here, we need to define a scalar function that we wish to differentiate, for which we choose
\begin{align}
    L_F = \sum_i \lVert \vec{F}_i \rVert \\
    L_\phi = \sum_i \lvert \phi_i \rvert
\end{align}
so that we can separately access the errors of force and potential VJP. For N-body simulations the potential sensitivities are usually not relevant, as the potential does not have a dynamical effect on particle positions, but accessing their accuracy may still be relevant for other convolution applications, for example the definition of loss functions (see section~\ref{sec:optimization}). We differentiate these losses with respect to all particle positions and masses and compute again relative errors with respect to a direct summation evaluation of the same quantities in Figure~\ref{fig:graderrors}. 

The gradients are clearly convergent in all cases. However, the most relevant combination $\frac{\partial L_F}{\partial \vec{x}}$ clearly converges slower than forces, only reaching $\lesssim 10^{-2}$ relative error at $p=5$. This is expected, since it is fundamentally a second-order quantity -- similar to the tidal field. In particular the $p=1$ case gives a consistent O(1) error, as the multipole expansion does not include second order terms in this case.

We conclude that the FMM implementation and the gradient evaluation have the expected convergence behaviour. It is worth noting that while the gradients deviate quantitatively slightly from the gradients that are obtained for direct summation, they are still exact gradients of the numerically evaluated FMM up to the aforementioned discontinuities.
\subsection{Performance break-down}

\begin{figure}
    \includegraphics[width=\columnwidth]{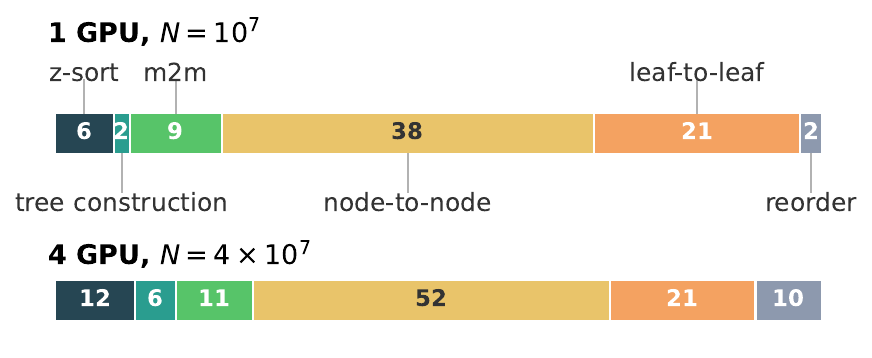}
    \caption{Execution time in ms for running all steps of the FMM ($p=5$) with forces returned in input order for a single GPU setup with $10^7$ points (top) and a 4-GPU setup with $4 \times 10^7$ points. The final reordering step can be avoided in most applications.}
    \label{fig:fmmsteps}
\end{figure}

We set up particles in a uniform random distribution and profile with default parameters (most importantly $p=5$ and $\theta_{\mathrm{max}} = 0.8$) the execution time of the individual steps of the FMM evaluation (1) for $N=10^7$ on a single GPU (2) for $N=4 \cdot 10^7$ particles on one full node with 4 GPUs. We exclude JIT-compilation time from these measurements and run each step $O(100)$ times in a loop to get accurate measurements. The corresponding measurements are shown in Figure~\ref{fig:fmmsteps}. Note that the total run-time of the full FMM may be slightly faster than the sum of these individual steps, because the compiled computation can eliminate the need to materialize some intermediate results \citep{jax_2018}.

We note that the single-GPU performance is dominated by the node-to-node (M2L) interactions and the leaf-to-leaf direct summations. Further, we note that sorting the particles in z-order is extremely performant and the tree-construction takes a negligible 2ms on top. This is slightly faster than the tree-construction time reported in \citet{stuecker_2026}, because we are not using any regularization here. This is in significant contrast to many N-body simulation codes that perform tree-construction on CPU, where tree construction can constitute a substantial cost \citep[e.g.][]{springel_2021,potter_2017}. The M2M translation has a minor, but clearly measurable contribution, which can largely be attributed to the high expansion order $p=5$ and a slightly less optimized kernel layout. The final reordering step brings forces back into input-order and can in principle be skipped if a simulation were to continue from the sorted output order.

The multi-GPU benchmark shows overall a relatively similar behaviour, but several steps are slightly more expensive due to the required steps to set up and perform communication -- in particular affecting the performance of tree construction, node-to-node translation and reordering.

\subsection{Scaling}

\begin{figure}
    \includegraphics[width=\columnwidth]{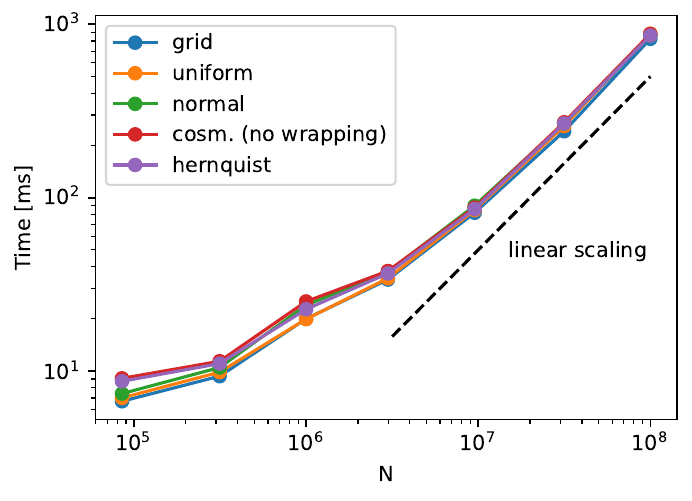}
    \caption{Performance and scaling of the FMM for different particle loads with default settings. Scaling is linear beyond $N \gtrsim 10^7$ and the method adapts very well to differently clustered problems.}
    \label{fig:performance_distributions}
\end{figure}

We aim to evaluate how well the presented algorithm generalizes to different problem setups. For this we test at different particle counts the performance of (1) a regular distribution on a cubic grid, (2) a uniform random distribution (3) a multivariate Gaussian distribution, (4) the final particle load of a cosmological simulation run with \textsc{disco-dj} \citep{List_2026} with box size chosen to yield a density of one particle per $h^{-3} \text{Mpc}^3$ and (5) a Hernquist sphere \citep{hernquist_1990}. Note that masses, length scales and softening don't have a performance impact so they are irrelevant here, but only the clustering of the distribution might change the tree and evaluation structure. We don't use periodic boundaries for any of the cases, so the obtained forces for the cosmological case are not meaningful, but the benchmark here should approximately represent the computational cost expected for such setups (within less than a factor 2 margin).

In Figure~\ref{fig:performance_distributions} we show the resulting execution times for a single GPU as a function of problem size. Encouragingly, the execution time is quite independent of the problem setup -- probably owed to the tree structure having by definition the same constant depth and similar particle counts per node for any setup. Further, we note that the algorithm scales linearly for $N \gtrsim 10^{7}$, as is expected for the FMM \citep{greengard_rokhlin_1987}. For $N \ll 10^{7}$ the scaling is sublinear, likely because the GPU is not fully saturated yet.

\begin{figure}
    \includegraphics[width=\columnwidth]{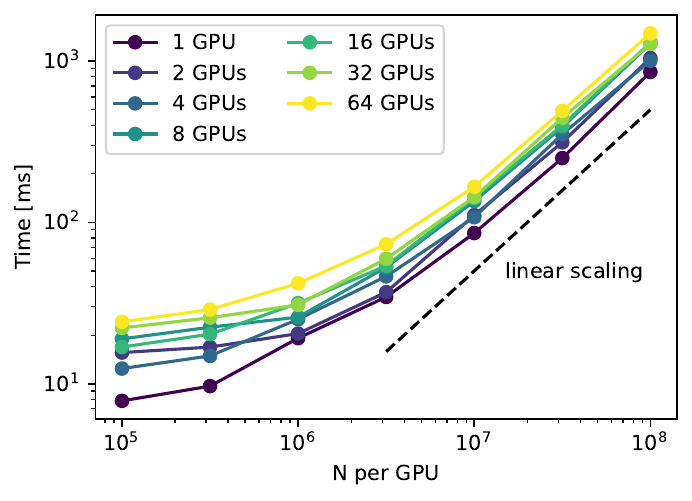}
    \caption{Performance and scaling of the FMM with device counts. In the GPU-saturating regime $N_{\mathrm{per\,GPU}} \gtrsim 10^7$ the efficiency decreases by less than a factor 2 when scaling from 1 to 64 devices.}
    \label{fig:performance_devices}
\end{figure}

Next we evaluate the scaling of the algorithm with the device count in Figure~\ref{fig:performance_devices}. For this we use a uniform random distribution and scale the particle count proportionally to the number of GPUs, so that e.g. the 64 GPU case with $N_{\text{per GPU}} = 10^8$ has a total problem size of $N = 6.4 \cdot 10^9$. We note that the linear scaling beyond $N_{\text{per GPU}} \gtrsim 10^7$ is nicely maintained for distributed setups. Further, we note that in the regime where GPUs are saturated, the decrease in efficiency from $1$ to $64$ GPUs is less than a factor 2. The main efficiency decreases are found when crossing from 1 to 2 GPUs (because communication becomes necessary) and when crossing from 4 GPUs to 8 GPUs (because internode communication is slower than intra-node communication). However, these measurements show that communication speed is not a major bottleneck for scaling to large device counts.

\subsection{Performance versus force accuracy}

We aim to compare the performance and force accuracy of the FMM implementation in \textsc{jz-fmm} with the ones in two widely used N-body simulation codes \textsc{gadget4} \citep{springel_2021} and \textsc{pkdgrav3} \citep{potter_2017}. \textsc{gadget4} and its predecessors are probably the most widely adopted cosmological simulation codes and have a long tradition in the cosmology community. \textsc{gadget4} uses the Message Passing Interface (MPI) to parallelize a pure CPU implementation of the FMM with a top-down-built octree structure and offers a large number of additional features, like hierarchical time-stepping, an optional acceleration based opening criterion, long range forces that can be calculated through a particle mesh, hydrodynamics and baryonic physics and many more. On the other hand \textsc{pkdgrav3} is a more recent N-body code that uses a hybrid CPU/GPU approach and also exhibits a notably similar feature set, but without baryonic physics. In \textsc{pkdgrav3} the tree and the node-node interactions of the FMM are handled on CPU, whereas the direct summation on the lowest level is accelerated through GPUs.

\begin{figure}
    \includegraphics[width=\columnwidth]{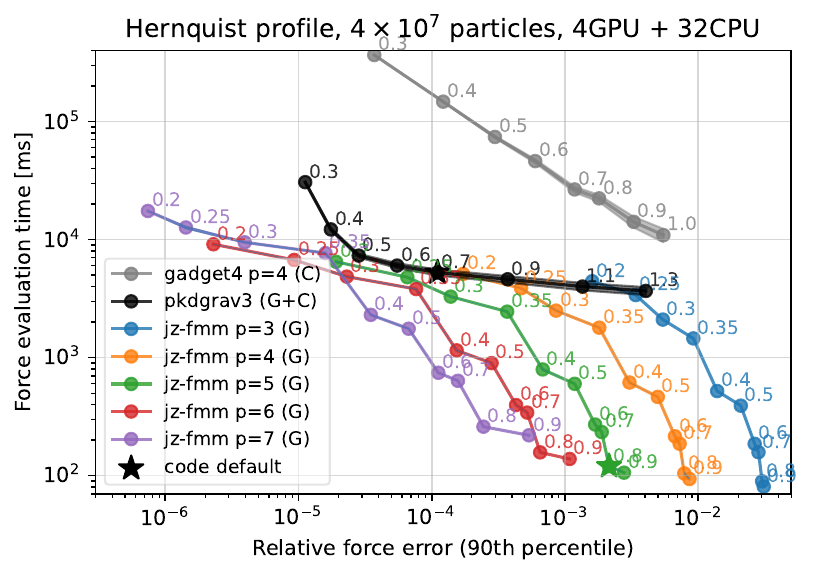}
    \caption{Comparison of the FMM in the force-accuracy versus evaluation time against \textsc{gadget4} and \textsc{pkdgrav3}. All benchmarks were run on a single node with 4 GPUs and a 32 core CPU -- not accurately mirroring the cost ratio of CPU versus GPU hardware. The numbers next to the points indicate the used opening angle. In \textsc{jz-fmm} choosing a higher expansion order is almost always preferable over decreasing the opening angle for achieving higher force accuracy. Results for double precision are shown in Figure~\ref{fig:performance_double}.}
    \label{fig:performance_codes}
\end{figure}

We choose these two codes as comparison points, because they are very mature, they represent well established standards in the community, and they allow us to contrast our pure GPU approach with a hybrid GPU/CPU and a pure CPU approach. However, it is worth noting that it is fundamentally difficult to perform a fair comparison of GPU and CPU codes, since hardware prices and energy requirements are significantly varying over time. Further, we only have access to the GPU nodes of the Leonardo booster partition, so that we will simply compare performance for a single of those nodes exhibiting 32 CPU cores and 4 GPUs. A more fair comparison would have to use at least a CPU-dedicated node with O(128-256) CPU cores as the base-line for the CPU code \textsc{gadget4}, so it is important to keep in mind that the performance of \textsc{gadget4} may appear in our test a factor 4-8 slower than in a fair comparison.

Another dimension of challenge for performance comparisons is that the codes use quite different choices for opening criterion, tree structure, softening, and the direct summation boundary. To compensate for this, we choose to compare performance in a two-dimensional space of force accuracy versus execution time. We do this as follows: We set up a load of $4 \cdot 10^7$ particles sampled from a Hernquist sphere, and we evaluate reference forces through direct summation with zero softening, Kahan summation and double precision. We then load the same particle distribution as initial conditions in each code and perform an integration for $O(100)$ time-steps with fixed time steps of negligible size $\Delta t \approx 0$ so that effectively the same distribution is evaluated multiple times. Periodic boundaries and adaptive time-stepping are turned off, all computations are done in 32bit floating point precision, the softening is set to zero, and we use the geometric opening criterion choice for all codes. For \textsc{gadget4} we use the $p=4$ expansion, since it appears to be the best option for the tested accuracy/performance domain, based on the figures in the paper. For \textsc{pkdgrav3}, we modify the value for the parameter nGroup to $256$, as officially recommended for execution with GPUs. We then average the reported total execution time of the domain-decomposition, tree-construction and force calculation, but exclude all other aspects (e.g. input/output, time integration). This way we get a good proxy for the time that is required for evaluating one full force computation with the FMM in all codes. Finally, we write out the forces and evaluate the per particle relative error with respect to the direct summation result and save the 90th percentile of the distribution. We repeat these steps, but with multiple different choices for the opening angle for each code. For \textsc{jz-fmm} we had to increase the allocation factors of the interaction list and the communication buffers for small opening angles, because the default allocation factors are targeted at the default opening angle of $\theta_{\mathrm{max}}=0.8$.

The resulting measurements are shown in Figure~\ref{fig:performance_codes}. First of all we note the accuracy and performance at default parameters: \textsc{pkdgrav3} has a small default relative force error of $\sim 10^{-4}$ with an execution time around 5 seconds. \textsc{jz-fmm} at $p=5$ has a notably larger default relative force error of $\sim 2 \cdot 10^{-3}$, but at a much lower execution time of $\sim 100$ milliseconds. The execution time in \textsc{pkdgrav3} scales very little with the opening angle, whereas it increases quite significantly in \textsc{jz-fmm}. On the other hand \textsc{gadget4} has a very regular and predictable scaling with the opening angle.

Comparing the different expansion orders in $p$ in \textsc{jz-fmm} shows that increasing the expansion order has a very minor effect on execution time, but a very significant effect on force accuracy. This trade-off seems to come out very different from CPU codes \citep{springel_2021} so that it is almost always beneficial to use a higher expansion order on GPU when force-accuracy is a concern. For $p=7$ with an opening angle of $0.6$ \textsc{jz-fmm} offers the same force accuracy as the default setup of \textsc{pkdgrav3}, but is almost an order of magnitude faster.

We conclude that the pure GPU implementation of the FMM in \textsc{jz-fmm} compares quite favourably with other state-of-the-art codes. It provides at all force-accuracies significantly improved performance. At the default opening angle and expansion order that lead to a relative force error of $\sim 2 \cdot 10^{-3}$, it is by notably more than an order of magnitude faster than both \textsc{pkdgrav3} and \textsc{gadget4}.

That said, \textsc{jz-fmm} does not yet offer a full drop-in replacement for such codes, as important performance relevant features like adaptive time-stepping and a long-range force-split are absent. However, it goes to show that dramatic performance improvements in N-body simulation codes can still be achieved when approaching the problem from a GPU native perspective.

\subsection{Integration Convergence}

\begin{figure}
    \includegraphics[width=\columnwidth]{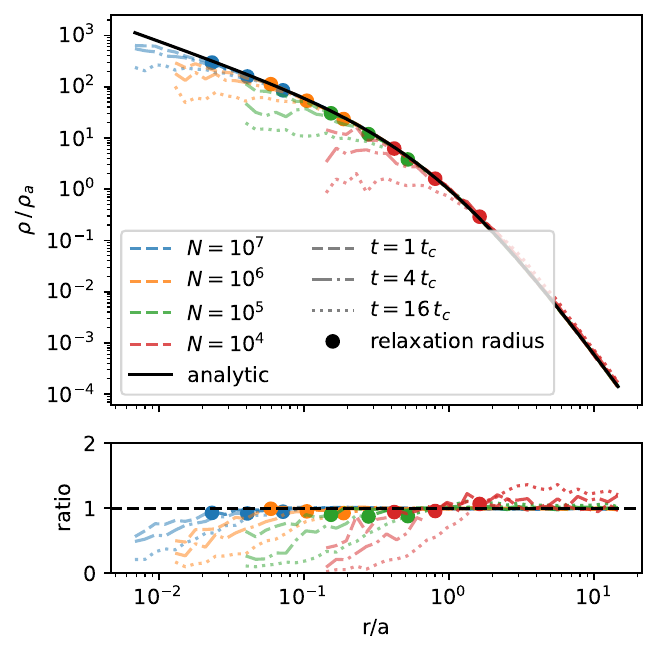}
    \caption{Profile convergence for an evolved Hernquist sphere (top) and residuals from the equilibrium expectation (bottom). The integration maintains the equilibrium system well beyond the relaxation radius.}
    \label{fig:hernquist_sim}
\end{figure}

To verify that the default force accuracy in \textsc{jz-fmm} is sufficient to reach convergence in N-body systems, we set up equilibrium Hernquist distributions with scale radius $a$ and evolve them for $1$, $4$, and $16$ circular orbit time-scales $t_c$ evaluated at $r=a$. We use a Plummer softening of $\epsilon = 0.002a$ and time-steps of size $0.01 t_c$. For a perfect Vlasov-Poisson system the Hernquist profile should not evolve at all, but the softening and the discreteness noise lead to deviations from the equilibrium state. Following the standard two-body relaxation scaling \citep{binney_tremaine_2008}, we adopt the local estimate
\begin{align}
    t_{\mathrm{rel}}(r) = \frac{0.1\,N(<r)}{\ln(r_{\mathrm{max}} / \epsilon)} \frac{t_{\mathrm{c}}(r)}{2},
    \qquad r_{\mathrm{max}} = 10^6 a,
\end{align}
where $N(<r)$ is the number of particles enclosed within $r$. Following the general convergence argument of \citet{power_2003}, we define the relaxation radius as the radius where the simulation time equals the relaxation time.
 
The resulting final profiles are shown in Figure~\ref{fig:hernquist_sim}. We can see expected deviations from the equilibrium profile below the relaxation radius to form core like structures. However, beyond the relaxation radius, the simulations converge reliably towards the correct solution.

\subsection{Simulation gradients}\label{sec:simgrad}
\begin{figure}
    \includegraphics[width=\columnwidth]{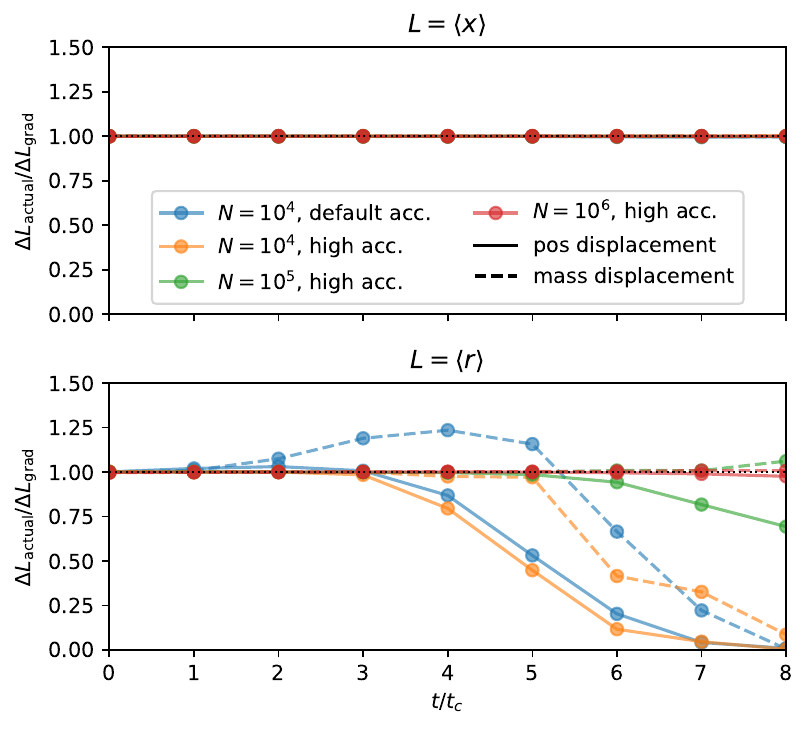}
    \caption{Accuracy tests for gradients of the centre of mass (top) and a structural quantity (bottom) of a Hernquist sphere evolved for varying amounts of time. A value of $\Delta L_{\mathrm{actual}} / \Delta L_{\mathrm{grad}} = 1$ means that gradients perfectly predict the response to modifications in the initial conditions.  Centre of mass gradients seem to be always perfect due to momentum conservation. Structural gradients are sensitive to integration time and numerical effects, but they converge with parameters that increase the accuracy of the simulation.}
    \label{fig:loss_experiment}
\end{figure}

We aim to evaluate how reliably gradients in \textsc{jz-fmm} are approximated for long N-body simulations. The core concern that needs to be evaluated is whether the piece-wise differentiability of the tree-based FMM could compromise gradient based methods, like e.g. a gradient descent or Hamiltonian Monte Carlo. The force field is discontinuous at node boundaries and the resulting jumps are not represented in the gradients. However, these jumps are expected to be rather small (compare Figure~\ref{fig:forceerrors}) and their amplitude can numerically be controlled for with the expansion order $p$ and the opening angle $\theta_{\mathrm{max}}$.

To evaluate this, we set up an N-body simulation of a Hernquist sphere that is truncated at $20a$ and simulate with a force softening of $\epsilon = 0.1a$ for varying times. At the final state we define two different scalar functions corresponding to the x-component of the centre of mass and the average radius
\begin{align}
    \langle x \rangle = \frac{\sum m_i x_i}{\sum m_i} \\
    \langle r \rangle = \frac{\sum m_i r_i}{\sum m_i}
\end{align}
We advect the gradients of these functions up to the initial conditions and then consider two types of modifications to the initial state, changing either the initial position or the mass of all particles:
\begin{align}
    \Delta \vec{x}_i &= 10^{-4} a \frac{\vec{g}_{\vec{x},i}}{\lVert \vec{g}_{\vec{x},i} \rVert} \\
    \text{or \quad} \Delta m_i &= 10^{-4} m_i \frac{g_{m,i}}{\lvert g_{m,i} \rvert}
\end{align}
where $\vec{g}_{\vec{x},i}$ is the gradient of the final scalar function with respect to the coordinates of the $i$th particle, and $g_{m,i}$ of its mass. Note that we have also tested velocity displacements, but they behave almost identically to position displacements (due to the quick mixing of position and velocity coordinates), so we will not discuss them separately here. If the gradients are correct, we may predict the change of the scalar function in the final state as
\begin{align}
    \Delta L &= \sum \vec{g}_{\vec{x},i} \cdot \Delta \vec{x}_i \\
    \text{or \quad} \Delta L &= \sum g_{m,i} \cdot \Delta m_i
\end{align}
We can compare this expected difference to the actual difference that we obtain when running a simulation from the perturbed initial state.

The results are displayed in Figure~\ref{fig:loss_experiment}, the top panel showing the results for the centre of mass scalar function and the bottom panel showing the results for the averaged radius scalar. Solid lines show the results for position displacements, whereas dashed lines show mass displacements. Different colours show variations with numerical accuracy parameters, such as the number of particles $N$, the opening angle $\theta$ and the expansion order $p$. The default accuracy case corresponds to $p=5, \theta_{\mathrm{max}}=0.8$, whereas the high accuracy case is $p=7, \theta_{\mathrm{max}}=0.4$.

There are several relevant observations to make:
\begin{itemize}
    \item The centre of mass case seems to be always described perfectly. This is not too surprising, since the mutual cell--cell formulation conserves momentum \citep{dehnen_2000,dehnen_2002} and centre of mass changes are correctly propagated even when numerical accuracy is poor.
    \item The radial loss case is notably more complex, showing clear deviations for longer simulations with a significant dependence on numerical parameters.
    \item At fixed numerical parameters, mass gradients tend to be more accurate than position gradients. As we have seen in Figure~\ref{fig:graderrors}, the FMM is more accurate for mass tangents than for position tangents. However, mass and position tangents mix over time in the integration, so that mass gradients are not fundamentally better behaved than position gradients, but rather they show a delayed onset of the inaccuracy.
    \item Numerical accuracy requirements for gradients seem to be significantly higher than for standard integration. E.g.~while $p=5, \theta=0.8$ may already be considered a fairly high force accuracy, we can clearly see gradient improvements when enhancing the force accuracy. Further, increasing the particle count from $N=10^4$ to $10^5$ has also a notable effect on the accuracy showing that both force accuracy and discreteness effects need to be controlled for.
\end{itemize}

We conclude that the gradient implementation in \textsc{jz-fmm} is generally convergent. Since centre of mass gradients seem to be perfect at any accuracy level, we may expect that in field level reconstruction scenarios it is easily achievable to get the overall positions and velocities of larger objects to converge -- independently of complicated internal dynamics. On the other hand, bringing the internal structure to convergence is probably also possible, but may need notably more numerical care.

\section{Satellite Reconstruction}

\begin{figure*}
    \includegraphics[width=\textwidth]{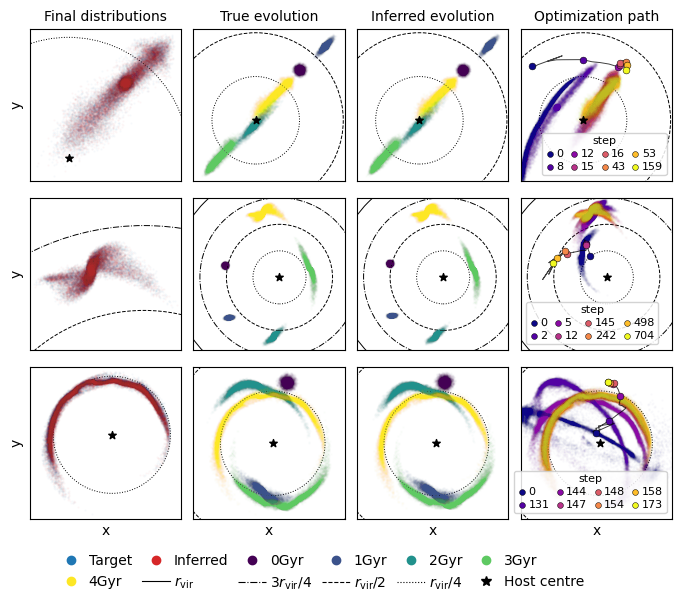}
    \caption{Examples of satellite reconstructions where the gradient descent found the global minimum of the loss. The left column shows the target and inferred distribution overplotted, whereas the second and third panels show their evolution separately. The fourth column shows the path of the initial centre of mass position and the final particle distribution through the gradient descent. Steps are chosen to display roughly equal distance in the logarithm of the loss.}
    \label{fig:success_reconst}
\end{figure*}

As a proof of concept of how differentiable simulations may be used to address deeply non-linear inference problems we consider the disruption of a satellite halo in a massive cluster environment. The primary goal here is to show how the numerical problem can be solved, whereas we leave the adaptation to realistic observational scenarios to future investigation.

\subsection{Target setup}

We consider a cluster with mass $M_\mathrm{host} = 10^{15} M_{\odot}$ and parameterize its analytical Hernquist host potential \citep{hernquist_1990} through a Navarro--Frenk--White (NFW)-inspired concentration \citep{navarro_1997} defined as $c_{\mathrm{host}} \equiv r_{\mathrm{vir}}/a_{\mathrm{host}}=6$ (corresponding to a virial radius $r_{\mathrm{vir}} = 2.1 \text{Mpc}$ and scale radius $a_{\mathrm{host}} \approx 351 \mathrm{kpc}$). We consider different evolved realizations of satellites as target distributions. We sample each satellite's orbit from the host profile's phase space distribution \citep{hernquist_1990} (excluding orbits with peri-centre below $0.1 r_{\mathrm{vir}}$ and apocentres above $r_{\mathrm{vir}}$) at $t = 0\text{Gyr}$ and choose Hernquist profiles with masses randomly sampled in $\log_{10} M \in 10...11$ and fixed concentration $c=8$, with the particle distribution truncated by excluding particle apocentres beyond 10 times the scale radius. To keep computational cost minimal, we use only $N = 2 \times 10^4$ particles here, but we evaluate the scaling with particle resolution later.

We consider 10 different initial satellite distributions by varying the seed of the random number generator. We evolve each satellite up to 5 different times $t=0,1,2,3,4\text{ Gyr}$ to create a total of 50 different target distributions. The simulations use 10 time steps per $\text{Gyr}$, a large softening of $4\textsc{kpc}$ (corresponding to $33-71\%$ of the satellite's scale radius, depending on the mass) and the default force accuracy parameters.

\subsection{Optimization setup} \label{sec:optimization}

We use the position of all final particles to define a loss function based on the maximum mean discrepancy (MMD) \citep{gretton_2012,sejdinovic_2013}:
\begin{align}
G_\epsilon(\boldsymbol{x}_i,\boldsymbol{x}_j)
    &= \sqrt{|\boldsymbol{x}_i-\boldsymbol{x}_j|^2+\epsilon^2},
    \label{eq:softened-distance-kernel}\\
\mathcal{L}_{\mathrm{MMD}}
    &= -\frac{1}{L_0}\sum_{i,j}w_iw_j
       G_\epsilon(\boldsymbol{x}_i,\boldsymbol{x}_j),
    \label{eq:mmd-loss}\\
(\boldsymbol{x}_i,w_i)
    &\in
    \left\{
        (\boldsymbol{x}_{\mathrm p,i},m_{\mathrm p,i} / M_p),
        (\boldsymbol{x}_{\mathrm t,i},-m_{\mathrm t,i} / M_t)
    \right\}.
    \nonumber
\end{align}
where $L_0 = 1\text{ Mpc}$ is used to obtain a dimensionless loss, subscripts 't' refer to target particles and 'p' to predicted ones, and we use a loss softening equal to the gravitational softening.  Normalizing each distribution by its total mass makes this loss insensitive to differences between $M_{\mathrm p}$ and $M_{\mathrm t}$ which will be kept fixed and identical.\footnote{In cases where they are not identical, their correspondence may be enforced through a separate loss term.} We prefer the softened distance kernel over other kernels here (e.g. a Plummer potential) to still obtain significant gradients at large distances. The loss corresponds to a convolution and can be evaluated efficiently through a single FMM calculation on the combined target and predicted particle positions with oppositely signed weights.


We then consider gradient descent optimizations of the initial subhalo position and velocity that are initialized with the same setups as the target, but different seeds. In particular, we also use a separate seed for the particle realization of the Hernquist profile so that a simulation with the correct position and velocity  still leads to a non-zero loss $L_{\text{ref}}$. Satellite mass and concentration are kept fixed to the true values. For the gradient descent we use a limited-memory Broyden--Fletcher--Goldfarb--Shanno (L-BFGS) optimizer \citep{liu_nocedal_1989} with memory size of 10 and a maximum of 5 line-search steps. We stop an optimization run when the loss does not improve over the previous minimum by more than a relative amount of $10^{-3}$ over 50 steps or when a total of 2000 steps is exceeded.

\subsection{Results}

To get a good overview of how reliably the optimization converges to the global minimum, we run per target 100 different initialization seeds. In Figure~\ref{fig:success_reconst} we show the lowest loss reconstructions that are obtained for a few examples of the 4Gyr targets. Clearly these are excellent reconstructions of the systems and they accurately describe the full history. However, it is important to note that the majority of gradient descents end up in local minima, which seem to be very abundant in the considered loss landscape. Some examples of such local minima are shown in Figure~\ref{fig:failed_reconst}. 

\begin{figure*}
    \includegraphics[width=\textwidth]{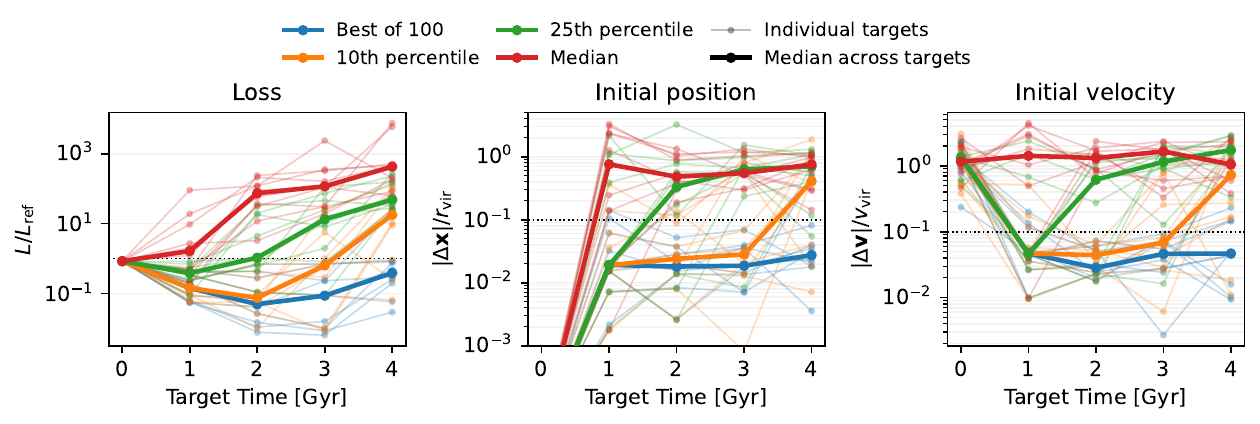}
    \caption{Analysis of the gradient descent convergence probability as a function of the length of the evolved target simulation. For each target seed the (loss-selected) percentiles of 100 initializations are shown as transparent lines, whereas the solid lines show the median across target seeds. The number of local minima increases significantly with evolution time and leads to more required initializations for reliable recovery of the global minimum. However, the global minimum of the loss reliably recovers the correct initial position and velocity of the satellite -- except for t=0 where the target particle positions contain zero information about the initial velocity.
    }
    \label{fig:convergence_grid}
\end{figure*}

To evaluate the convergence probability quantitatively we show in Figure~\ref{fig:convergence_grid} the distribution of losses and the recovery of the initial position and initial velocity. Importantly, the percentiles are always selected based on the loss, so that e.g. the best loss cases are very reliably recovering the initial position and velocities, without knowledge beyond the final particle distribution. The one notable exception is $t=0$ where the target particle distribution does not contain any information about the orbital velocity yet. Interestingly, the gravitational evolution allows breaking this degeneracy at later times.

Further, we note that the probability per run to converge to a local (rather than the global) minimum, increases with the evolution time. For example for $t = 1 \text{Gyr}$ more than a quarter of the initial seeds find the global minimum, whereas at $t = 4 \text{Gyr}$ less than 10\%. Probably local minima associated with orbits that reach the same final position from a different direction and velocity become more frequent and widely spread in the initial space the more orbits a satellite has gone through.

\begin{table}
    \centering
    \caption{Computational cost of a simulation based gradient descent as a function of the
    number of particles $N$ averaged over 10 different target systems, run on a single A100 GPU. For small $N$ scaling is sublinear, since the GPU resources are not fully utilized. Each evaluation corresponds to a simulation + gradient backward integration with $40$ time-steps and default force accuracy parameters.}
    \label{tab:reconstruction_cost}
    \begin{tabular}{rccc}
        \hline
        $N$ & Evaluations & Total time [min] & Time/eval. [s] \\
        \hline
        $2\times10^{4}$ &
        $566 \pm 205$ & $7.0 \pm 2.4$ & $0.76 \pm 0.05$ \\
        $10^{5}$ &
        $604 \pm 276$ & $12.6 \pm 5.4$ & $1.27 \pm 0.07$ \\
        $10^{6}$ &
        $483 \pm 206$ & $39.5 \pm 15.9$ & $5.0 \pm 0.7$ \\
        $10^{7}$ &
        $509 \pm 269$ & $195.5 \pm 97.5$ & $23.5 \pm 3.9$ \\
        \hline
    \end{tabular}
\end{table}

To provide a rough estimate of the costs of such gradient descent reconstructions we select, for each of the 10 target systems, the initialization that reaches the lowest loss after the gradient descent. We rerun these initializations (which generally take longer to converge than poor reconstructions that end in local minima) at varying particle resolutions $N$ and list their run-time in Table~\ref{tab:reconstruction_cost}.
We note that the scaling is sublinear for $N \ll 10^7$ as expected from the benchmarks in Figure~\ref{fig:performance_devices}, since low particle counts cannot fully utilize a GPU's resources. A rough estimate of the time per gradient evaluation is
\begin{align}
    t_{\mathrm{eval}} \approx 3 \cdot n_{\mathrm{steps}} \cdot t_{\mathrm{fmm}}
\end{align}
where $t_{\mathrm{fmm}}$ is the time required for a single force-evaluation. The factor $3$ accounts for effectively $3$ force evaluations being needed per time-step -- one for the forward integration, one for the backward integration and one for the gradient advection. Assuming $t_{\mathrm{fmm}} \sim 100\text{ ms}$, this gives about $12$ seconds for the $10^7$ particle case with $40$ time-steps, rather than the measured $23.5$ seconds. Contributions to this difference include the final loss calculation and its gradient, as well as the VJP of the force calculation being slightly more expensive than a pure force calculation, because the near-field particle-particle interactions require more expensive dipole interactions. For this particular setup we also found in some time-steps very large tree nodes that would interact with almost every other node -- owed to the large empty volume outside the stripped region. In an additional test we turned on tree regularization \citep{stuecker_2026}, which improved performance of the $10^7$ particle gradient evaluation by up to 35\%. However, the performance gains from regularization depend on the particle distribution, so we leave it disabled by default.

\subsection{Summary}

We summarize our main findings:
\begin{itemize}
    \item The gradient implementation in \textsc{jz-fmm} functions correctly and allows solving reconstruction problems with algorithms that utilize gradients.
    \item The cost of gradient evaluations is quite manageable so that reconstructions may already be obtained on the time scales of minutes or hours with a single GPU for sufficiently small setups.
    \item The tidal tails of the stripped satellite allow to uniquely reconstruct the initial position and velocity of the satellite. The global minimum of the loss is close to the optimal reconstruction in all evolved distributions. This is a noteworthy result, since pure information of the final object's centre of mass position would be insufficient for such a reconstruction.
    \item The considered loss landscape is plagued by local minima. The number of local minima grows significantly with evolution time. Thus, a reliable recovery of the global minimum requires multiple initializations or usage of more explorative techniques, such as Hamiltonian Monte Carlo \citep{neal_2011} or tempered sampling \citep{marinari_parisi_1992}.
\end{itemize}

It is worth noting that the reconstruction here is only possible, because of a strong prior on the initial particle distribution (limited to follow a halo profile). Without such a prior, final particle positions alone generally underconstrain the initial phase-space state of a full N-body system. The core strength of differentiable simulations is to build the bridge between an evolved final distribution and a strong initial prior -- an approach that has already been applied in cosmological field-level reconstruction simulations to great success \citep[e.g.][]{jasche_wandelt_2013,mcalpine_2026}.

\section{Conclusions}

Here, we have presented a novel code \textsc{jz-fmm} for evaluating and differentiating gravitational forces of N-body systems through the FMM. We have shown that gradient advection through the FMM can primarily be evaluated through a standard FMM with modified input multipoles. \textsc{jz-fmm} was designed from scratch with a pure GPU based computation model, and we have shown that it significantly outperforms the force evaluation of other CPU based or hybrid N-body codes.

Beyond this, the new code shines through numerous desirable features such as
\begin{itemize}
    \item A simple \textsc{pip} based installation and a convenient \textsc{python} interface allowing to easily compose components of the code in user-defined ways.
    \item Generality in dimension and expansion order.
    \item Easy extensibility, e.g. through custom external potentials or through new convolution kernels.
    \item Bit-perfect reproducibility of the FMM.
    \item Bit-perfect backwards integration (when using the integer lattice integrator).
    \item Compatibility with \textsc{JAX}'s just-in-time (\textsc{JIT}) compilation.
    \item Gradient evaluation for particle positions, velocities and masses.
\end{itemize}

We have shown for a simplified satellite reconstruction problem that the differentiable simulations can be used to solve complicated optimization problems with moderate computational costs. However, the loss landscape of such deeply non-linear problems is highly non-trivial, and it may require sophisticated optimization techniques to guarantee the recovery of a global minimum. Further, we have seen that gradients converge well with numerical accuracy parameters, but reliable gradient computations require more conservative parameter choices than standard N-body simulations.

The code as it stands is already well suited for applications in N-body simulations with isolated boundary conditions (both when gradients are needed and when not). However, the full potential of the new method will be unlocked when combining the differentiable tree algorithm with cosmological initial- and boundary conditions, for example by incorporating it through a short-range long-range force split with mesh based cosmological simulation codes like \textsc{DISCO-DJ} \citep{List_2026}. This will make it possible to advance cosmological reconstruction techniques to significantly smaller scales with an efficient and flexible short-range force calculation.

\section*{Acknowledgements}

The authors thank Adrian Gutierrez Adame for helping to set up benchmarks of \textsc{pkdgrav3} and Oliver Hahn and all members of the Vienna Cosmology group for helpful discussions. This research was funded in whole or in part by the Austrian Science Fund (FWF) [10.55776/ESP705]. The authors acknowledge access to the EuroHPC supercomputer LEONARDO, hosted by CINECA (Italy) through the AURELEO call.

\subsection*{Use of artificial intelligence (AI) tools}

The algorithm and principal structure of the code were developed by the authors with minor assistance from OpenAI GPT models. Later stages of code development benefited substantially from these tools for template generalization, micro-optimizations, testing, documentation and packaging. The manuscript was written primarily by the authors, with assistance from the same tools for language polishing, consistency checks and initial drafts of some paragraphs. The authors take full responsibility for the code, scientific results and final manuscript.

\section*{Data Availability}

The \textsc{jz-fmm} source code, together with test and benchmark scripts, is publicly available under the MIT licence at \url{https://github.com/jstuecker/jzfmm}. The package is also available through PyPI at \url{https://pypi.org/project/jzfmm/}. Documentation, installation instructions and example simulations are available at \url{https://jstuecker.github.io/jzfmm/}.



\bibliographystyle{mnras}
\bibliography{jzfmm} 




\appendix

\section{Gradients in N-body systems} \label{app:nbodyjvpvjp}
In this Appendix we derive two interesting symmetries between the VJP and JVP in the N-body system. We did not use these symmetries in the implementation of \textsc{jz-fmm}, because the symmetries only apply to position and velocity tangents, but not to mass tangents. However, we list them here, since they may prove useful at some point in the future, e.g. if position/velocity based forward tangents are needed.

\subsection{Force symmetry}

The force field has the Jacobian
\begin{align}
    \nabla_{x_k} \vec{F}_i &= -G \sum_{j \neq i} \nabla_{x_k} \nabla_{x_i} g(\vec{x}_i - \vec{x}_j) m_j \nonumber \\
    &= \sum_{j \neq i}  \delta_{ik} \mat{T}_{ij} m_j - \delta_{jk} \mat{T}_{ij} m_j \nonumber \\
    &= \delta_{ik} \sum_{j \neq i} m_j   \mat{T}_{ij} - (1 - \delta_{ik}) m_k \mat{T}_{ik} \nonumber
\end{align}
where $\mat{T}_{ij} = -G\nabla \nabla g(\vec{x}_i - \vec{x}_j)$ is the tidal tensor per unit source mass. We find the symmetry
\begin{align}
    m_i \nabla_{x_k} \vec{F}_i &= m_k \nabla_{x_i} \vec{F}_k
\end{align}
Therefore, for the equal mass case $m_i = m_k$ the Jacobian is symmetric and the VJP and JVP of the force field computation are identical.

\subsection{Symplectic JVP and VJP relation}

A symplectic matrix $\mat{J}$ has the properties
\begin{align}
    \mat{J}^T \mat{\Omega} \mat{J} &= \mat{\Omega} \\
    \mat{J}^{-1} &= \mat{\Omega}^{-1} \mat{J}^T \mat{\Omega} \\
    \mat{\Omega} &= \begin{pmatrix} 0 & \mat{I}_n \\ -\mat{I}_n & 0 \end{pmatrix}
\end{align}
Assuming equal-mass particles, the Jacobian of all final positions and velocities $\begin{pmatrix} \vec{x}^N & \vec{v}^N \end{pmatrix}^T$ with respect to the initial ones $\begin{pmatrix} \vec{x}_0^N & \vec{v}_0^N \end{pmatrix}^T$ is a symplectic matrix. Therefore, we can relate 
\begin{align}
    \frac{\partial J}{\partial (\vec{x}, \vec{v})_{0}} &= \begin{pmatrix}\vec{g}_{x}^N \\ \vec{g}_{v}^N\end{pmatrix}^T \mat{J} \nonumber \\
    &= \begin{pmatrix}\vec{g}_{x}^N \\ \vec{g}_{v}^N\end{pmatrix}^T (\mat{\Omega} \mat{J}^{-1} \mat{\Omega}^{-1})^T \nonumber \\
    \frac{\partial J}{\partial (\vec{x}, \vec{v})_0}^T &= \mat{\Omega} \mat{J}^{-1} \mat{\Omega}^{-1} \begin{pmatrix}\vec{g}_{x}^N \\ \vec{g}_{v}^N\end{pmatrix}
\end{align}
Therefore, a VJP through the forward integration can be evaluated as a JVP through the inverse integration, after rotating the input and output by the symplectic matrix -- and vice versa.
Therefore, the VJP implementation in \textsc{jz-fmm} allows in principle also to evaluate forward derivatives.

\section{Additional results}

\begin{figure}
    \includegraphics[width=\columnwidth]{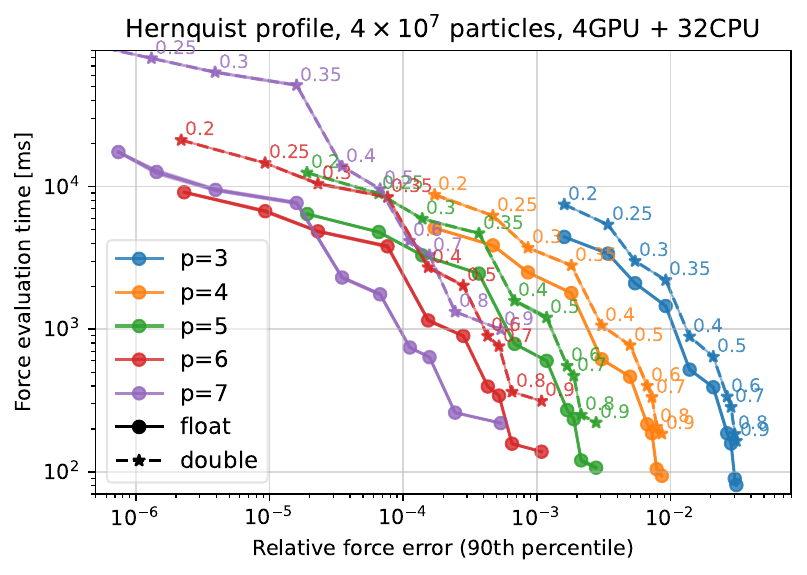}
    \caption{Same as Figure~\ref{fig:performance_codes}, but comparing the single versus double precision evaluation in \textsc{jz-fmm}. For $p \lesssim 5$ double precision requires approximately twice the evaluation time of single precision, whereas for $p > 5$ the performance gap becomes bigger, because the intermediate terms become too large to be kept in efficient GPU register space (also known as register spilling).}
    \label{fig:performance_double}
\end{figure}

\begin{figure*}
    \includegraphics[width=\textwidth]{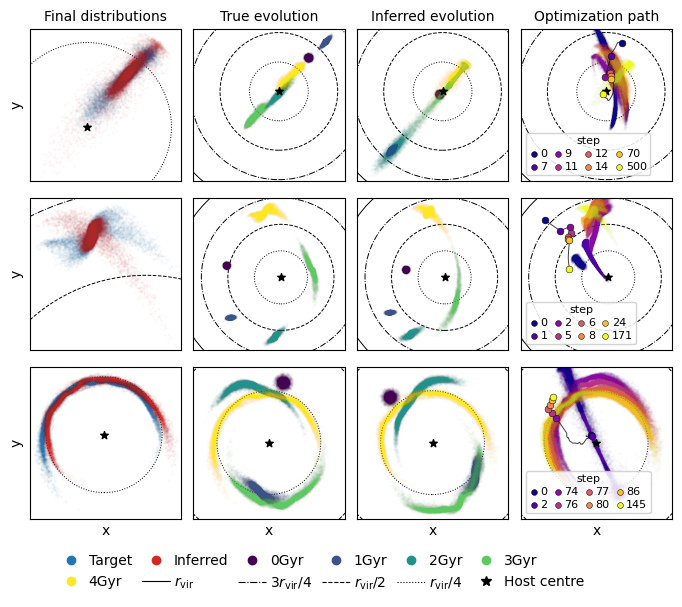}
    \caption{Examples of gradient-descent reconstructions that converged to a local minimum of the loss function. Each row shows the same target seeds as in Figure~\ref{fig:success_reconst}, but with a gradient descent that converged to a local minimum solution. Row 1 shows an example of a descent towards the wrong initial side of the host, row 2 a solution that seems qualitatively on a similar orbit to the correct solution and row 3 a reconstructed orbit that approached the target from the opposite direction. While all these cases recover part of the target distribution, none of them is remotely close to the global minima from Figure~\ref{fig:success_reconst} in terms of loss.}
    \label{fig:failed_reconst}
\end{figure*}


\bsp	
\label{lastpage}
\end{document}